\documentstyle[aps,amssymb,psfig]{revtex}
\begin{document}

\author{Simonetta Frittelli$^a$ \and Thomas P. Kling$^b$ \and Ezra T.
Newman$^c$}

\address{
$^a$Department of Physics, Duquesne University,
Pittsburgh, PA 15282\\
$^b$ Department of Physics, Astronomy and Geophysics,
Connecticut College, New London, CT 06320\\
$^c$Department of Physics and Astronomy,
University of Pittsburgh,
Pittsburgh, PA 15260
}
\title{
\rightline{\small{\em To appear in Phys. Rev. D\/}}
Fermat Potentials for Non-Perturbative Gravitational Lensing}
\date{April 12, 2002}
\maketitle

\begin{abstract}

The images of many distant galaxies are displaced, distorted and often
multiplied by the presence of foreground massive galaxies near the line of
sight; the foreground galaxies act as gravitational lenses. Commonly, the lens
equation, which relates the placement and distortion of the images to the real
source position in the thin-lens scenario, is obtained by extremizing the time
of arrival among all the null paths from the source to the observer (Fermat's
principle). We show that the construction of envelopes of certain families of
null surfaces consitutes an alternative variational principle or version of
Fermat's principle that leads naturally to a lens equation in a generic
spacetime with any given metric. We illustrate the construction by deriving the
lens equation for static asymptotically flat thin lens spacetimes. As an
application of the approach, we find the bending angle for moving thin lenses
in terms of the bending angle for the same deflector at rest. Finally we apply
this construction to cosmological spacetimes (FRW) by using the fact they are
all conformally related to Minkowski space.

\end{abstract}

\section{Introduction}


It is the purpose of this note to point out, study and apply an idealized
construction of gravitational lens equations that is of potential use in
many physical situations -- from exact lensing to the weak-field thin-lens
scenario -- by means of an alternative version of the standard usage of
Fermat's principle.

The fundamental aspect of gravitational lensing theory is the construction
of the past light-cone of an observer. This directly leads to the idea of
the mapping from the space of images -- the celestial sphere of the observer
-- to the space of the sources -- usually to the ``source plane'', though
this specialization is by no means necessary. The mapping is carried out by
following, backwards in time, the null geodesics of the lightcone, from the
observer to the source. In other words, by knowing where an image appears on
the observer's celestial sphere and knowing the null geodesics that generate
the past null cone, one could in principle follow the rays back to the
source. In addition, it is often of considerable importance to know the
transit times between the emission of light and its arrival at the observer.
In fact, in view of Fermat's principle, the actual path taken by light is a
local extremum of the transit time of all possible neighboring null paths,
which leads to the formulation of gravitational lensing via Fermat's
principle. Much of contemporary lensing theory is based on the construction,
on a simple background (either Minkowski or a cosmological spacetime), for
weak fields and with a thin-lens and small-angle approximation, of an
appropriate transit time function. Then, by the local extremization of the
time function, a lens equation is constructed. In the usual approach, the
time function (referred to as a Fermat potential) represents the transit
time along all possible null curves, not necessarily geodesic, that connect
the source and the observer\cite{EFS,Perlick,Petters}. The extremization
produces a suitable lens equation by selecting those curves that are
geodesic, i.e., the variation of the travel time with respect to the paths,
restricted to the condition that the paths be null, is equivalent to the
geodesic equation with null tangents.

We wish to show that there is an attractive alternative (and significantly
different) variational principle (an alternative Fermat's principle) for use
in lensing that can be applied, at least in principle (and, with
approximations, also in practice), in a generic situation. The basic
framework is to begin with a general four-dimensional Lorentzian spacetime
where the geometry (the metric) of the spacetime is to be considered as the
``gravitational lens''. In other words, the detailed lens properties are to
be coded directly into the metric tensor. We then consider a 2-parameter
family of null (or characteristic) surfaces passing through an observer's
world line at a given time. In fact, we assume that the family of surfaces
is sufficiently generic for the null normals at the observer to be distinct
and span the sphere of null directions (often just an open neighborhood of
the sphere is sufficient). This family of null surfaces then contains all
the points on the observer's lightcone (or the open neighborhood), since at
the intersection of the observer's worldline with each surface the normal
vector to the surface is null, geodesic, and lies on the surface. Each null
geodesic that passes through the observer's world line on each of the null
surfaces can then be followed into the past. These are the rays that an
observer sees and constitute his \textit{celestial\/} sphere (or an open
neighborhood of it).

Now consider a point source of light moving along some given (source) timelike
world line. We are interested in those null geodesics moving back in time from
the observer to the source, i.e., those geodesics that the observer ``sees'' as
coming from the source. At \textit{any given observer moment\/}, the observer
will ``see'' a number of different light rays (or images), each of which, in
general, will have intersected the source (or equivalently, will have been
emitted by the source) at different source times. These countably few null
geodesics lie on countably few of the null surfaces in the family that we are
considering. Thus there are a number of surfaces in the 2-parameter family that
intersect the worldline of the source at a point that can be connected to the
observer by a null geodesic on the surface. However, at \textit{any observer
moment,\/} all (or almost all) the other surfaces in the family intersect the
source's world line as well, at varying times. In general, however, there are
no curves lying on these surfaces connecting the source intersection point to
the observer, that are null geodesics. In fact, most of the curves that can be
used to connect the source and the observer on each of these null surfaces are
piecewise spacelike and/or null. For this reason, we prefer to drop any
reference to paths, and keep the argument in terms of the null surfaces. The
\textit{source time\/} $t$ at the intersection point is thus a function on the
\textit{sphere\/} of null surfaces intersecting the observer's world-line at a
particular observer's time $T_0$:

\begin{equation}\label{Ts}
   t=T(\theta,\varphi,T_0) ,  
\end{equation}

\noindent where $(\theta,\varphi)$ label the null surfaces.
We now ask for the local extremes of $T$ as a function of $(\theta,\varphi)$:

\begin{mathletters}\label{DT}
\begin{eqnarray}
     \partial_\theta T 
    &\equiv &T,_\theta(\theta,\varphi,T_0) = 0,			 \\
     \partial_\varphi T 
    &\equiv &T,_\varphi(\theta,\varphi,T_0)=0.
\end{eqnarray}
\end{mathletters}

\noindent This operation picks out those null surfaces that possess curves
from the observer traveling backwards to the source {\it which are null
geodesics\/}. It constitutes our version of Fermat's Principle.
This version differs from the usual one not merely in form. If one thinks of
it in terms of paths connecting the source and the observer, this version of
Fermat's principle allows for curves that are neither null nor geodesic. It
concentrates on and varies the null surfaces rather than the null curves.

For this reason we think of our version of Fermat's principle as an
alternative to the usual one. Correspondingly, we refer to the function $T$
as a {\it generalized Fermat potential.\/} With respect to the underlying
meaning of our version of Fermat's principle, one can see that
Eqs.~(\ref{DT}) are equivalent to the construction of the envelope of the
null surfaces passing through the observer, which in turn is the past
lightcone of the observer. Thus we arrive at the observer's lightcone
starting from surfaces in a way analogous to the usual approach, which
arrives at the observer's lightcone starting from null paths.

In Section II we describe this construction in greater detail and justify the
claim that it does pick out the null surfaces so that null geodesics connect
the observer with the source thus yielding the past lightcone. In Sec.~III we
illustrate the construction for the trivial case of Minkowski space without a
lens, while in Sec.~IV we illustrate it for a Minkowski space background with a
static thin lens, the conventional scenario. In Sec.~V we apply the
construction to thin lenses that are moving in order to obtain the corrections
to the lens equation due to the motion of the lens. In Section~VI these ideas
are applied to the FRW cosmologies using the fact that they all are conformally
related to Minkowski space. We conclude, in Section~VII, with remarks and an
outlook of the potential applications of the construction.

We find this method quite enlightening concerning the underlying ideas in
gravitational lensing theory. And although it might be difficult to apply in
many practical physical situations without the use of heavy approximations,
preliminary calculations suggest that, with further development, it could
well be of practical use.

\section{Implicitly defined generalized Fermat potentials}


We assume that we have a four-dimensional Lorentzian manifold, $(M,x^a)$
with a given Lorentzian metric, $g_{ab}$. We further assume that null
geodesics and null surfaces (solutions to the eikonal equation) can be
constructed, in some local coordinates, wherever needed. In general, null
surfaces develop wavefront singularities and, in principle, one must worry
about them. However, the trouble caused by the singularities is easily
bypassed if the surfaces are parametrized by the null geodesics that rule
them. The fundamental objects for us will be {\it complete integrals of
the eikonal equation:\/} two-parameter families of foliations of the
spacetime by null surfaces, so that at any spacetime point there is a
sphere's worth of null surfaces through that point. The two parameters are
arbitrary, but for our purposes we specify them as points on the sphere. In
the following, then, we adopt this particular choice of parameters to label
the solutions of the eikonal equation in the complete integral, and we
specify them by complex stereographic coordinates $(\zeta,\bar\zeta)$,
rather than the regular spherical coordinates $(\theta,\varphi)$. The
complete integral is given analytically by an expression of the form

\begin{equation}  \label{Z}
	Z(x^a,\zeta,\bar\zeta)
\end{equation}

\noindent such that for each fixed value of $(\zeta,\bar\zeta)$ the level
surfaces of (\ref{Z}) are null. Equivalently, (\ref{Z}) satisfies the eikonal
equation

\begin{equation}  \label{Eikonal}
g^{ab}\partial_a Z\partial_bZ =0,
\end{equation}

\noindent with two further conditions: the four functions  $(Z,\partial_\zeta
Z,\partial_{\bar\zeta}Z,\partial_{\zeta\bar\zeta}Z)$  form a rank-four set
with respect to $x^a$ {\it almost everywhere}, whereas {\it globally}, the
three functions $(Z,\partial_\zeta Z,\partial_{\bar\zeta}Z)$ form a
rank-three set. How these conditions are used is explained shortly below. In
addition, the null covectors $l_a(x^a,\zeta,\bar\zeta)\equiv \partial_a 
Z(x^a,\zeta,\bar\zeta)$ at fixed $x^a$ sweep out all null directions as
$(\zeta,\bar\zeta)$ vary, thus reproducing the local lightcone. We rewrite
(\ref{Z}) as the two-point function

\begin{equation}  \label{Gted}
       G(x^a,x_0^a,\zeta,\bar\zeta) 
\equiv Z(x^a,\zeta,\bar\zeta)
      -Z(x_0^a,\zeta,\bar\zeta) =0.
\end{equation}

\noindent Any two points $(x^a,x_0^a)$ satisfying Eq.~(\ref{Gted}) for fixed 
$(\zeta,\bar\zeta)$ lie on the same null surface. It is clear that the
gradient of $G$ with respect to either point is a null covector. If $x_0^a$
is fixed, however, then the points $x^a$ that satisfy Eq.~(\ref{Gted}) for
each $(\zeta,\bar\zeta)$ lie on different null surfaces, but $x_0^a$ belongs
to all. So we think of $x_0^a$ as a special point where all the null
surfaces intersect.

The point $x_0^a$ is chosen to lie on the observer's worldline given by
$x_0^a=x_0^a(T_0)$, $T_0$ being some observer time. Though it is not needed
and is used only for convenience, we choose, in the neighborhood of the
source, a special coordinate system where the source remains at a fixed
spatial point, $x^i$, and only the source time $x^0=t$ evolves. The equation 
$G(x^a,x_0^a,\zeta,\bar\zeta)=0$ then becomes

\begin{equation}  \label{G1ted}
	G(x^i,t,x_0^a(T_0),\zeta,\bar\zeta)=0
\end{equation}

\noindent or, rewritten (since $\partial_t G\neq 0$ by assumption)
as

\begin{equation}  \label{Ghat}
	t =T(x^i,T_0,\zeta,\bar\zeta),
\end{equation}

\noindent it gives rise to our time function, Eq.~(\ref{Ts}). As a matter
of notation, by analogy with the usual theory of gravitational lensing it is
natural for us to refer to $T$ as a {\it generalized Fermat potential.\/}
However, the function $G$ arises more naturally from the starting point of
the eikonal equation, therefore, with a slight abuse of terminology, we
refer to it as the {\it implicit Fermat potential,\/} and for all
practical purposes we use Eq.~(\ref{G1ted}) instead of Eq.~(\ref{Ghat}).

The extremal values of $T$ are calculated in terms of the implicit Fermat
potential $G$ by implicit differentiation of Eq.~(\ref{G1ted}). Thus
$\partial_\zeta T$ and $\partial_{\bar\zeta}T$ are obtained from

\begin{mathletters}\label{minimal}
\begin{eqnarray}
G,_t \; \partial_\zeta T +\partial_\zeta G &=&0,  \label{min1} \\
G,_t \; \partial_{\bar\zeta}T +\partial_{\bar\zeta}G &=&0,  \label{min2}
\end{eqnarray}
\end{mathletters}

\noindent so that the vanishing of $\partial_\zeta T$ and
$\partial_{\bar\zeta}T$ implies the vanishing of both $\partial_\zeta G$
and $\partial_{\bar\zeta}G$, and the reverse, i.e.,

\begin{equation}  \label{Equivalent}
	(\partial_\zeta T,\partial_{\bar\zeta}T)=0 
\Leftrightarrow 
	(\partial_\zeta G,\partial_{\bar\zeta}G)=0.
\end{equation}

\noindent Thus, setting $\partial_\zeta G=\partial_{\bar\zeta}G=0$ gives us
the extreme values of $T$. There is a deeper meaning to the extremization of 
$T$, in connection with the null surfaces in the family, as we subsequently
show.

Returning to Eq.(\ref{Gted}), with fixed $x_0^a$, we see that we have a
two-parameter family of surfaces through a fixed point $x_0^a$. A new
surface, with common tangent curves all passing through $x_0^a$, is the
envelope of the family and is constructed by requiring that $\partial_\zeta G
= \partial_{\bar\zeta}G = 0$. The triple

\begin{mathletters}\label{triple}
\begin{eqnarray}
	G(x^a,x_0^a,\zeta,\bar\zeta) &=&0,  \label{triple1} \\
		    \partial_\zeta G &=&0,  \label{triple2} \\
		\partial_{\bar\zeta}G&=&0,  \label{triple3}
\end{eqnarray}
\end{mathletters}

\noindent defines the envelope. For $x^a\neq x_0^a$ and in a region 
{\it without\/} wavefront singularities\cite{E.N.,Gilb}, Eqs.~(\ref{triple2})
and (\ref{triple3}) can be algebraically inverted so that

\begin{mathletters}\label{gamma}
\begin{eqnarray}
	    \zeta &=&\Upsilon (x^a,x_0^a), \\
	\bar\zeta &=&\bar{\Upsilon}(x^a,x_0^a).
\end{eqnarray}
\end{mathletters}

\noindent When they are substituted into (\ref{triple}), we have the
equation for a three-surface, namely the envelope:

\begin{equation}  \label{envelopeted}
	G_{env}(x^a,x_0^a) 
 \equiv G(x^a,x_0^a,\Upsilon(x^a,x_0^a),\bar\Upsilon(x^a,x_0^a)) 
      = 0.
\end{equation}

\noindent That $G_{env}(x^a,x_0^a)=0$ is a null surface follows from both the
fact that $\partial G/\partial x^a$ is a null vector (by assumption) and from
(\ref{triple2})-(\ref{triple3}). That it is the null cone of the point
$x_0^a$ is shown by demonstrating that at $x^a=x_0^a$ the surface has a
sphere's worth of tangents. This is seen by the following argument: in the
limit as $x^a\rightarrow x_0^a$, Eqs.~(\ref{triple2})-(\ref{triple3}) are
identically satisfied for all values of $(\zeta ,\bar\zeta)$. This is a
degenerate point where Eqs.(\ref {gamma}) \textit{do not\/} hold. It then
follows that the tangent vector, given by

\begin{equation}  \label{lightcone}
   \left.\frac{\partial G_{env}}{\partial x^a}\right|_{x=x_0} 
  =\left.\frac{\partial}
	      {\partial x^a}G(x^a,x_0^a,\zeta,\bar\zeta)\right|_{x=x_0},
\end{equation}

\noindent has multiple values, yielding a null vector that sweeps out the
tangent-space null cone at $x_0^a$, as $(\zeta,\bar\zeta)$ moves over
the sphere.

Our task is now to show that this extremization process picks out the
particular surfaces that connect the source and observer by null geodesics;
namely, the values of $(\zeta,\bar\zeta)$ for each pair  $(x^a,x_0^a)$ given
explicitly by (\ref{gamma}) -- or implicitly by
(\ref{triple2})-(\ref{triple3}) -- label a geodesic null vector whose
integral line passes through $x_0^a$ and $x^a$. In order to show this, we
notice that when Eqs.~(\ref{triple2})-(\ref {triple3}) can not be inverted,
i.e., when the Jacobian

\begin{equation}  \label{J}
   J =\left| 
	\begin{array}{ll}
		\partial_{\zeta\zeta}G \hspace{0.4cm} 	& 
		\partial_{\bar\zeta\zeta}G 	\\ 
							&
		  				\\ 
		\partial_{\zeta\bar\zeta}G 	   	&   	
		\partial_{\bar\zeta\bar\zeta}G
	\end{array}
      \right|
\end{equation}

\noindent vanishes, the null surfaces develop wavefront
singularities\cite{E.N.}. But by the assumption that
$Z(x_0^a,x^a,\zeta,\bar\zeta)$ is a complete integral and the rank conditions,
in either case the three equations (\ref{triple}) can be solved for three (say
$x^i$) of the four spacetime coordinates $x^a$, in terms of a fourth one
($x^*$) and $(\zeta,\bar\zeta)$, namely:

\begin{equation}  \label{parametric}
	x^i = X^i(x_0^a,x^*,\zeta,\bar\zeta).
\end{equation}

\noindent As a consequence of (\ref{triple}), it turns out that the curves
described by (\ref{parametric}) by keeping $(\zeta,\bar\zeta)$ fixed are null
geodesics. This can be seen by implicit differentiation of (\ref{triple})
with respect to $x^*$. Defining $t^a\equiv (1,\partial X^i/\partial
x^*|_{\zeta,\bar\zeta})$, we obtain

\begin{mathletters}\label{triple*}
\begin{eqnarray}
	G,_a(x^a,x_0^a,\zeta ,\bar\zeta)t^a &=&0,  \label{triple*1} \\
	             \partial_\zeta G,_at^a &=&0,  \label{triple*2} \\
	        \partial_{\bar\zeta}G,_at^a &=&0  \label{triple*3}
\end{eqnarray}
\end{mathletters}

\noindent where ($G,_a,\partial_\zeta G,_a,\partial_{\bar\zeta}G,_a)$ are
independent,  by the rank condition, again. Eq.~(\ref{triple*1}) implies that
$ t^a$ has no component pointing out of the surface, so $t^a=\alpha
g^{ab}(G,_b+\beta\partial_\zeta G,_b+\bar\beta\partial_{\bar\zeta}G,_b)$.
Using this in Eqs.~(\ref{triple*2}) and (\ref{triple*3}) implies that $\beta
=\bar\beta=0$ and hence $t^a=\alpha g^{ab}G,_b$, and thusthe tangent vector
is proportional to $g^{ab}G,_b$ being therefore null, and the integral curves
are null geodesics.

We thus obtain the parametric description of all the null geodesics through
$x_0^a$. Notice that having this parametric description of the lightcone (in
terms of the observer's celestial sphere) is entirely equivalent to having a
lens equation relating the angular position of the source at a given distance
to the angular position of the image on the celestial sphere, plus a time of
arrival equation yielding the transit time of the light signal from the
source to the observer\cite{UnivLens,EFN}. To see this, the point $x_0^a$ is
taken to represent a point on the observer's worldline. By treating
$(\zeta,\bar\zeta)$ as the celestial sphere of the observer, $x^*$ as a
measure of ``distance'' backwards along the null geodesic to a source and
$x^i$ as representing the ``time'' and ``angular position'' of the source at
the moment of emission, we have constructed a lens equation together with the
transit time equation. Thus one can see that, in a completely general
context, our version of Fermat's Principle, Eqs.~(\ref{DT}), leads
immediately to the construction of the observer's past lightcone, with
associated lens and time of arrival equations.

When we extremize the function $T$ we could obtain a maximum, a minimum or a
saddle point. We wish to know how to rephrase the conditions for maximum,
minimum or saddle in terms of our implicit Fermat potential, G. In principle,
we need to look at the eigenvalues of the matrix containing the second
derivatives of $T$ evaluated at the extrema. We now switch to real variables
$(u,v)$ instead of $(\zeta,\bar\zeta)$, via $\zeta \equiv u+iv$ , and calculate
the eigenvalues of the matrix

\begin{equation}
	\delta^2 T 
 \equiv \left( 
	\begin{array}{ll}
	T,_{uu} \hspace{0.2cm} 	& T,_{uv} \\ 
			  	&  	  \\ 
	T,_{uv} 		& T,_{vv}
	\end{array}
	\right).
\end{equation}

\noindent Because of $T$ being implicitly defined via (\ref{G1ted}), we just
need to take second derivatives of (\ref{G1ted}). For instance, if we take
two $u-$derivatives of (\ref{G1ted}) (with (\ref{Ghat}) for $t$) we obtain

\begin{equation}
	  G,_{uu} 
	+ T,_u\left( 2 G,_{ut}+ G,_{tt} T,_u\right) 
 	+ G,_t T,_{uu} =0.
\end{equation}

\noindent Since we are interested only in the value of $T,_{uu}$ at the
extremum, then $T,_u=0$ and

\begin{equation}
	T,_{uu} = -\frac{G,_{uu}}{G,_t},
\end{equation}

\noindent which presents no difficulty, since by assumption $G,_t$ must be
nonvanishing. Likewise, the other components of the matrix come out
proportional to the second derivatives of $G$:

\begin{equation}  \label{eq:19}
	\delta^2 T 
     = -\frac{1}{G,_t} 
		\left( 
		\begin{array}{ll}
		G,_{uu} \hspace{0.2cm} 	& G,_{uv} 	\\ 
					&  		\\ 
		G,_{uv} 		& G,_{vv}
		\end{array}
		\right).
\end{equation}

\noindent Thus, up to a factor of $-1/G,_t$ the eigenvalues are the same as
those of $\delta^2G$. In order to fix unnecessary sign freedom (since $-G$
for our purposes is just as good as $G$), we assume that $G,_t<0$ at the
extremum in question. It remains to rephrase our conditions for maximum,
minimum and saddle in terms of the complex variable $\zeta$. Since 
$\zeta=u+iv$, we have $\partial/\partial u = \partial/\partial\zeta +
\partial/\partial\bar\zeta$ and $\partial/\partial v =
i(\partial/\partial\zeta - \partial/\partial\bar\zeta)$, thus

\begin{mathletters}\label{eq:20's}
\begin{eqnarray}
	G,_{uu} &=& G,_{\zeta\zeta} 
		  + G,_{\bar\zeta\bar\zeta} 
		  + 2G,_{\zeta\bar\zeta},	\label{eq:20} \\
	G,_{vv} &=& - G,_{\zeta\zeta} 
		    - G,_{\bar\zeta\bar\zeta} 
		    +2G,_{\zeta\bar\zeta} ,  	\label{eq:21} \\
	G,_{uv} &=& i( G,_{\zeta\zeta} 
		     - G,_{\bar\zeta\bar\zeta}).  \label{eq:22}
\end{eqnarray}
\end{mathletters}

\noindent If we think of $G$ as spin-weight zero function on the sphere
then, using the envelope condition, $G,_\zeta=G,_{\bar\zeta}=0$ we have

\begin{mathletters}\label{eq:23's}
\begin{eqnarray}
    \eth^2 G &=&(1+\zeta \bar\zeta)^2G,_{\zeta\zeta },  	 \\
\bar\eth^2 G &=&(1+\zeta \bar\zeta)^2G,_{\bar\zeta\bar\zeta}.\label{24}
\end{eqnarray}
\end{mathletters}

\noindent and hence

\begin{equation}
    \delta^2 T 
 = -\frac{2}{(1+\zeta\bar\zeta)^2G,_t} 
	\left( 
	\begin{array}{ll}
	\eth\bar\eth G+Re(\eth^2 G)\hspace{0.3cm} & -Im(\eth^2 G) \\ 
						  &  		  \\ 
	-Im(\eth^2 G) & \eth\bar\eth G-Re(\eth^2 G)
	\end{array}
	\right).
\end{equation}

\noindent The eigenvalue equation for $\delta^2 T$ is given by

\begin{equation}
   (\tilde{\lambda}-\eth\bar\eth G)^2-\eth^2 G\bar\eth^2G = 0,
\end{equation}

\noindent where $\tilde{\lambda} =-(1+\zeta\bar\zeta)^2G,_t\lambda/2$ and 
$\lambda$ is an eigenvalue of $\delta^2 T$. Since $G,_t$ is assumed to be
negative, then the sign of $\lambda$ is the same as the sign of
$\tilde{\lambda}$. The solutions are

\begin{equation}
    \tilde{\lambda}_{\pm }=\eth\bar\eth G\pm |\eth^2G|.
\end{equation}

\noindent Finally, the conditions are:

\begin{itemize}
\item if $|\eth\bar\eth G|<|\eth^2G|$, the extremum is a saddle, 
since the eigenvalues have opposite signs.

\item  if $\eth\bar\eth G>|\eth^2G|$, the extremum is a minimum,
since both eigenvalues are positive.

\item  if $\eth\bar\eth G<-|\eth^2G|$, the extremum is a maximum, since 
both eigenvalues are negative.
\end{itemize}

\noindent On the other hand, the envelope develops singularities when

\begin{equation}
\eth\bar\eth G^2=\eth^2G\bar\eth^2G
\end{equation}

\noindent as anticipated earlier (the vanishing of the determinant of the
Jacobian matrix). The type of extremum is of fundamental importance. For
instance, in a very weak field, no singularities develop in the observer's
lightcone, there is no multiplicity of images and the lens equation represents
a local minimum of the Fermat potential. Increasing the strength of the field,
in a situation where there are three images, one will be a minimum, one a
maximum and the remaining one will be a saddle. The minimum corresponds to the
primary image, produced by a ray that does not encounter a caustic in its path.
The saddle corresponds to a ray that passes the caustic once. The maximum
yields the ray that passes the caustic twice and yields the faintest image. The
generalization of this interpretation to cases of more than three images is
complicated\cite{EFS,Petters} and lies beyond our present interest.

\subsection{Parametric Version of Implicit Fermat Potential}

We finish this section by presenting a very important method for describing
implicit Fermat potentials. In many cases it is difficult or impossible (as we
will see) to obtain a closed-form $G$ for a family of surfaces
$G(x^a,x_0^a,\zeta,\bar\zeta)=0$, especially if the surfaces self-intersect and
have singularities. In such cases, it is simpler to describe the family of
surfaces in parametric form adapted to the null geodesics ruling the surface,
namely, by specifying the map

\begin{equation}
x^a=\Gamma^a(s,r,q,\zeta,\bar\zeta)  \label{parametricG}
\end{equation}

\noindent where $(s,r,q)\equiv s^i$ are coordinates on the surface, given for
each fixed value of $(\zeta,\bar\zeta)$. Since each member of the family (for
fixed $(\zeta,\bar\zeta)$) is a null surface, it is automatically ruled by null
geodesics (except at singular points of the surface). Therefore the surface
coordinates $s^{i}$ can always be chosen as geodesic coordinates: $(r,q)$
labeling the null geodesics ruling the surface, and $s$ as an affine parameter
along the geodesics. However, our interest is in surfaces that intersect at a
given point $x_0^a$. Therefore, all the surfaces in the family must contain the
common point  $x_0^a$. We can make use of this fact by adapting the coordinates
$(s,r,q)$ to this point. Since $s$ is a parameter along the geodesics, we fix
it by demanding that the 2-surface $s=0$ be transverse (to the null geodesics
in the null surface) and contain the point $x_0^a$. This is, then, the
initial-data surface for the geodesics in the null surface. On this
initial-data 2-surface, the origin of coordinates for $r$ and $q$ can be taken
to be $x_0^a$. In this manner, all the null surfaces in the family will single
out the observer's point as the point with $s=r=q=0$, namely:

\begin{equation}
x_0^a=\Gamma^a(0,0,0,\zeta,\bar\zeta)
\end{equation}

\noindent so that $\Gamma^a$ depends on $x_0^a$, i.e., 

\begin{equation}
x^a=\Gamma^a(s,r,q,x_0^a,\zeta,\bar\zeta).
\end{equation}

\noindent Often we take $x_0^a$ as the observation time
$T_0$ at the spatial origin and omit it from the equations.

In the case that the family of surfaces $G(x^a,x_0^a,\zeta,\bar\zeta)=0$ is
given parametrically via (\ref{parametricG}), the envelope construction
proceeds in an implicit manner as follows. In the first place, the ruling of
the null surface via null geodesics guarantees that the Jacobian matrix of the
map (\ref{parametricG}), namely $\partial\Gamma^a/\partial s^i$, has rank three
(except at singular points). Therefore, locally, from the equations
(\ref{parametricG}) one can always chose 3 equations that can be solved for
$s^i$. The fourth equation remains a function of $s^i$. So equations
(\ref{parametricG}) can always be viewed in the form

\begin{mathletters}\label{implicitGs}
\begin{eqnarray}
  \Gamma^i(s^i,\zeta,\bar\zeta;x_0^a)-x^i &=&0, \label{implicits} \\
    \Gamma(s^i,\zeta,\bar\zeta;x_0^a)-x^4 &=&0,  \label{implicitG} \\
                                   \Gamma &\equiv &\Gamma^4
\end{eqnarray}
\end{mathletters}

\noindent where (\ref{implicitG}) is interpreted as the equation
$G(x^a,x_0^a,\zeta,\bar\zeta)=0$ if $s^i$ is thought of as implicitly given in
terms of $x^i$ via (\ref{implicits}). In other words,

\begin{equation}\label{GImplicit}
	  G(x^a,x_0^a,\zeta,\bar\zeta)
\equiv \Gamma(s^i(x^i,\zeta,\bar\zeta;x_0^a),
		      \zeta,\bar\zeta;x_0^a)-x^4.  
\end{equation}

\noindent In the envelope construction we then have

\begin{equation}
       0=\left. \partial_\zeta G\right|_{x^a}
	=\left. \partial_\zeta\Gamma\right|_{s^i}
	+\left. \frac{\partial\Gamma}{\partial s^i}\right|_\zeta
         \left. \partial_\zeta s^i\right|_{x^i},  \label{DGImplicit}
\end{equation}

\noindent where $\partial_\zeta s^i|_{x^i}$ is determined by implicit
differentiation of (\ref{implicits}). Both Eqs.(\ref{GImplicit}) and
(\ref{DGImplicit}) will be extensively used in the following sections.

This appears to be a convenient method for constructing implicit Fermat
potentials and for obtaining lens equations, as demonstrated in the
following sections.

\section{The implicit Fermat potential in Minkowski spacetime}


The case of Minkowski spacetime is quite trivial but is helpful as an
illustration of the main ideas underlying the concept of the implicit Fermat
potential $G$. In cartesian coordinates $x^a$, null surfaces in Minkowski
spacetime are level surfaces of the solutions to the eikonal equation 
$\eta^{ab}G,_aG,_b=0$, with $\eta_{ab}=$diag$(1,-1,-1,-1)$. There is a
2-parameter family of distinct foliations by null planes with the two
parameters corresponding to a sphere's worth of null directions at any given
spacetime point. Using complex stereographic coordinates, the null directions
can be specified by $l^a(\zeta,\bar\zeta)$ given by

\begin{equation}\label{ell}
	l^a(\zeta,\bar\zeta)
   \equiv 
	\frac{1}{\sqrt{2}}\left( 
			1,\, \frac{(\zeta+\bar\zeta)}{1+\zeta\bar\zeta},
			  \,-\frac{i(\zeta-\bar\zeta)}{1+\zeta\bar\zeta},
			  \,-\frac{1-\zeta\bar\zeta}{1+\zeta\bar\zeta}\right).
\end{equation}

\noindent The null planes with null normal $l^a$ containing the observer's
location $x_0^a(T_0)$ are given by

\begin{equation}\label{G'}
	G(x^a,x_0^a,\zeta,\bar\zeta)
   \equiv 
	(x^a-x_0^a)l_a(\zeta,\bar\zeta)
   = 0.  
\end{equation}

\noindent The envelope of this 2-parameter family of null planes is the
lightcone of the point $x_0^a$ and is obtained by taking the partial
derivatives with respect to $\zeta$ and $\bar\zeta$ of the function 
$G(x^a,x_0^a,\zeta,\bar\zeta)$ and setting these derivatives equal to zero. In
this case:

\begin{mathletters}\label{Gderiv}
\begin{eqnarray}
	\frac{\partial G}{\partial\zeta } 
     &=&(x^a-x_0^a)\frac{\partial l_a}{\partial\zeta}=0,  \label{G1} \\
	\frac{\partial G}{\partial\bar\zeta} 
     &=&(x^a-x_0^a)\frac{\partial l_a}{\partial\bar\zeta}=0.  \label{G2}
\end{eqnarray}
\end{mathletters}

\noindent Eqs.~(\ref{G'}) and (\ref{Gderiv}) define a 3-surface in the
spacetime. One way to obtain a single equation defining such a surface is to
solve (\ref{Gderiv}) for $\zeta$ as a function of $x^a$, wherever possible, and
substitute $\zeta(x^a)$ back into (\ref{G'}). Then the envelope is thus given,
in principle, by

\begin{equation}
G_{env}(x^a)\equiv G(x^a,x_0^a,\zeta(x^a),\bar\zeta(x^a))=0
\end{equation}

In this case, this procedure can be carried out in closed form. Using the
notation $\Delta x^a\equiv x^a-x_0^a$, Eq.~(\ref{G2}) is equivalent to

\begin{equation}\label{G2implies}
(\Delta x - i\Delta y)\zeta^2 -2\Delta z\zeta -\Delta x -i\Delta y = 0
\end{equation}

\noindent with solution

\begin{mathletters}
\begin{eqnarray}
        \zeta(x^a) &=&\frac{\Delta z\pm \Delta r}
                           {\Delta x-i\Delta y}, \label{zeta} \\
    \Delta r &\equiv &\sqrt{(\Delta x)^2+(\Delta y)^2+(\Delta z)^2}.
\end{eqnarray}
\end{mathletters}

\noindent (The other equation, (\ref{G1}), is the complex conjugate of
(\ref{G2}) and yields no new information). Substituting $\zeta(x^a)$ given by
Eq.~(\ref{zeta}) and its complex conjugate $\bar\zeta(x^a)$ into Eq.(\ref{G'})
yields

\begin{equation}\label{flatcone}
 G_{env}(x^a,x_0^a)
=G(x^a,x_0^a,\zeta(x^a),\bar\zeta(x^a))
=\frac{1}{\sqrt{2}}(\Delta t\mp \Delta r)
=0,  
\end{equation}

\noindent where the minus sign is for the future lightcone and the plus sign
for the past lightcone of the point $x_0^a$. Consequently, the set of three
equations (\ref{G2implies}) and (\ref{G'}) are equivalent to the lens and time
of arrival equations. Because there is no time occurrence in (\ref{G2implies}),
then one can think of it as the lens equation since it relates the spatial
position of source $(x,y,z)$ to its corresponding image on the celestial sphere
$(\zeta,\bar\zeta)$. In principle, solving for two of $(x,y,z)$ in terms of
$(\zeta,\bar\zeta)$ and substituting into Eq.~(\ref {G'}) turns Eq.~(\ref{G'})
into an equation relating the time of emission $t$ from a source at $(x,y,z)$
with image direction $(\zeta,\bar\zeta)$ seen at the observer $x_0^a$.

\section{The implicit Fermat potential for a static asymptotically flat
thin-lens spacetime}


In geometrical terms, a thin lens can be modeled as a spacetime with
vanishing curvature everywhere except at a timelike surface, where the
``lens'' lives. Accordingly, the spacetime consists of two flat spacetimes,
matched appropriately at the lens' worldsheet, with the observer lying in
one flat spacetime and the source lying in the other one. We refer to these
two flat spacetimes as the observer's side and the source's side of the
spacetime.

It is clear that the families of null surfaces to be used on the observer's
side of the spacetime are all the null planes that contain the observer's
location, labeled by the direction of the null normal. On the source's side,
however, the appropriate null surfaces to use are not null planes but the
distortion of the null planes by the lens. For this reason they must, in
general, be given in parametric form. We use the parametric method, via
Eq.~(\ref{parametricG}), explained in Section II, for the envelope
construction.

For the purpose of clarity of presentation, we first develop our construction
of the null surfaces in a lower-dimensional setting. We then generalize the
argument to 3+1 dimensions.

\subsection{2+1 static lens planes}

The $2+1$ lensing setting is described in cartesian coordinates $x^a=(t,x,y)$.
The observer is at $x_0^a=(T_0,0,0)$, and the lens surface is a plane located
at $y=y_l$. The metric of the spacetime is flat everywhere except at the lens
plane:

\begin{equation}
ds^2=c^2dt^2-dx^2-dy^2.
\end{equation}

\noindent Our aim is to construct a 1-parameter family of null foliations which
in principle can be given as the level surfaces of a function  $G(x^a,\varphi)$
which replaces the $G(x^a,\zeta,\overline{\zeta})$ of the four dimensional
lensing scenario. (The origin, $x_0^a$, is omitted.) This function will be
entirely equivalent to the Fermat potential in this lensing scenario, and the
envelope of the surfaces $G(x^a,\varphi)=0$ will simultaneously give us the
lens equation and time of arrival. In the approximation of small angles, the
Fermat potential and the lens equation will reduce to the standard
astrophysical lensing scenario.

Roughly our construction is as follows; we define a 1-parameter family of null
surfaces, (parametrized by $\varphi$) by imagining a straight line of photons
traveling (backwards in time) parallel to each other (null rays) in a direction
making an angle $\varphi$ with the $y-$axis. As the photons move, they trace a
null plane in the spacetime. The moment each photon arrives at the lens plane,
it is ``detained'' for a length of time  ${\cal T}$ that depends only on the
point at which the photon hits the lens plane. After this time, the photon is
released in a new direction $\varphi' =\varphi-\alpha$, deviating by an angle
$\alpha$ from its original direction. The bending angle $\alpha$ depends only
on the point at which the photon hits the lens plane, and it should be entirely
determined by the ``detention time'' ${\cal T}$. Because different photons
arrive at the lens plane at different times and are detained for different
lengths of time, the wavefront they make up after leaving the lens plane is not
necessarily plane.

The scheme that we have just described has a corresponding analog in the
wavefront distortion produced by a thin sheet of glass with a variable index of
refraction. A lightray impinging nearly normally on the sheet will bend inside
the glass due to the component of the gradient of the index of refraction
perpendicular to its path, exiting the glass with a finite deviation angle. A
plane incident wavefront then comes out of the glass distorted. The local index
of refraction at the point of incidence multiplied by the thickness of the
glass is a measure of the time spent by each lightray inside the glass, or
equivalently, the ``detention'' time of light inside the glass. For later
purposes, it is important that we make following point with respect to the
limit in which the thickness of the glass tends to zero. The ``detention time''
in the region occupied by the glass defined in this way has two contributions:
the transit time in vacuum through the region (the ``vacuum'' time:
$c^{-1}\times $(thickness of the glass)) plus the amount of time that light is
delayed by the slowing-down in the presence of the medium, with respect to
vacuum. In the limit in which the thickness of the glass tends to zero, the
first (vacuum) term vanishes and the ``detention time'' coincides with the
time delay with respect to vacuum.

In the case of interest to us, the detention time ${\cal T}$ is determined by
the gravitational field of a deflector on the lens plane and it coincides with
the gravitational time delay with respect to vacuum:

\begin{equation}\label{tau}
	{\cal T}=\frac{-2}{c^3}\int Udl  
\end{equation}

\noindent where $U$ is the Newtonian potential of an isolated mass
distribution and the integration takes place along the null ray or line of
sight. The situation we are contemplating is the limiting case in which the
support of $U$ tends to zero, while the integral of $U$ remains constant,
namely: $U$ is a distribution along each null ray.

\begin{figure}
\centerline{\psfig{figure=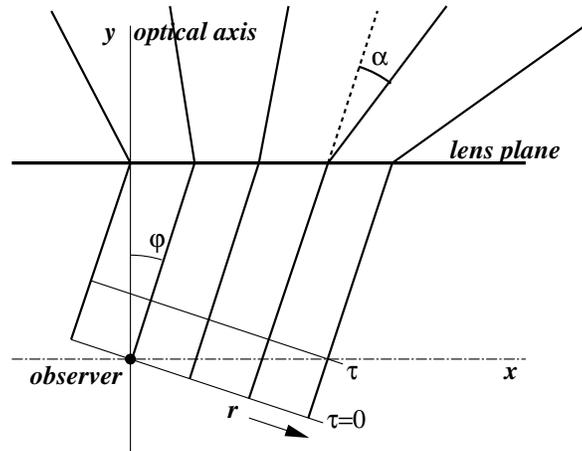,width=3in,angle=0}}

\caption{Parametric construction of the null surfaces.  On the observer side,
the surface is a null plane ruled by parallel lightrays making an angle
$\varphi$ with the $y-$axis.  The lightrays are labeled by their distance $r$
away from the observer, and are parametrized by the distance $c\tau$ away from
the wavefront that passes by the observer. On the other side of the lens, the
lightrays bend through an unspecified angle $\alpha$ that depends on the
position at which the lightrays reach the lens plane.  }

\label{fig:0}
\end{figure}

More precisely, {\it on the observer side} of the spacetime we consider the
parallel lightrays traced by a plane wavefront, moving (backwards in time) at
an angle $\varphi$ with the optical axis (the $y$-axis) as defining our null
surface. See Fig.~\ref{fig:0}. The wavefront passes by the observer at time $T_0$.
The (perpendicular) distance from the observer to the wavefront at later times
divided by $c$, which is denoted by $\tau$, can be used as a parameter along
each lightray. The individual lightrays at $\tau=0$, can be labeled by their
distance $r$ from the observer. In terms of these parameters, the time
coordinate of the lightrays is given by

\begin{equation}  \label{ttau}
	t = T_0-\tau,
\end{equation}

\noindent whereas the space coordinates of the lightrays are given by
 
\begin{mathletters}\label{fermato}
\begin{eqnarray}
	x &=&c\tau \sin \varphi +r\cos \varphi ,  \label{fermatox} \\
	y &=&c\tau \cos \varphi -r\sin \varphi .  \label{fermatoy}
\end{eqnarray}
\end{mathletters}

\noindent (Note that as $\tau $ increases time is going backwards.) It is clear
that $r=\tau=0$ defines the observer at $(x,y)=(0,0)$. Each lightray (for fixed
$\varphi$) reaches the lens plane at $\tau =\tau_l$ when $y=y_l$, hitting a
lens point $x=x_l$:

\begin{mathletters}\label{pretaulxl}
\begin{eqnarray}
	x_l &=&c\tau_l\sin \varphi +r\cos \varphi ,  \label{prexl} \\
	y_l &=&c\tau_l\cos \varphi -r\sin \varphi .  \label{pretaul}
\end{eqnarray}
\end{mathletters}

\noindent These, in turn, can be solved for $\tau_l$ and $r$ and yield

\begin{mathletters}\label{taulxl}
\begin{eqnarray}
	c\tau_l &=& x_l\sin\varphi +y_l\cos\varphi ,  \label{taul} \\
              r &=& x_l\cos\varphi -y_l\sin\varphi .  \label{s}
\end{eqnarray}
\end{mathletters}

\noindent Eqs.~(\ref{fermato}) hold while $\tau <\tau_l$.

On the other side of the lens line, the wavefronts will be determined by two
things: the bending angle $\alpha(x_l)$ and the gravitational time delay
imposed on the lightrays at the lens line, which we here denote by 
${\cal~T}(x_l)$. The bending angle is to be determined in terms of
${\cal~T}(x_l)$. The bending angle and the time delay can be incorporated into
the parametric equations for these lightrays in the following way:

\begin{mathletters}\label{fermats}
\begin{eqnarray}
	x &=& c(\tau-\tau_l-{\cal T})\sin(\varphi-\alpha) 
	      + x_l,  					\label{fermatsx}\\
	y &=& c(\tau-\tau_l-{\cal T})\cos(\varphi-\alpha) 
	      + y_l.  					\label{fermatsy}
\end{eqnarray}
\end{mathletters}

\noindent These equations hold for $\tau>\tau_l+{\cal T}$. For times in
between, the lightrays are ``delayed'' at the lens plane, so we have

\begin{mathletters}\label{fermatl}
\begin{eqnarray}
	x &=&x_l,  		\label{fermatlx} \\
	y &=&y_l,  		\label{fermatly}
\end{eqnarray}
\end{mathletters}

\noindent for $\tau_l<\tau <\tau_l+{\cal T}$. Eqs.(\ref{fermato}),
(\ref{fermats}), and (\ref{fermatl}) represent the parametric expression of a
one-parameter family of null surfaces which in principle can also be expressed
by a single equation of the form $G(x,y,t,\varphi)=0$ away from caustics.

On the observer side this can be done explicitly by eliminating $r$ from
Eqs.~(\ref{fermato}) yielding

\begin{equation}\label{xala}
        G(x,y,\tau,\varphi)
 \equiv y\cos\varphi +x\tan\varphi -\tau 
      = 0.
\end{equation}

\noindent On the source side, however, this can not be done explicitly. The
parameter that labels the null geodesics is $x_l$, as opposed to $r$ on the
observer side. They are related by Eq.~(\ref{s}). The idea would be to
eliminate $x_l$ between Eqs.~(\ref{fermatsx}) and (\ref{fermatsy}) but this is
impossible since both ${\cal T}$ and $\alpha $ are functions of $x_l$. We thus
need to define $x_l$ as a function of $x$ implicitly by (\ref{fermatsx}). We
take (\ref{fermatsy}) as the equation that defines the function $G$ where $x_l$
is the implicit function of $x$ given by (\ref{fermatsx}):

\begin{eqnarray}\label{xalasource}
	G(x,y,\tau,\varphi) 
   &=&  y 
      - y_l
      -(c\tau-x_l\sin\varphi-y_l\cos\varphi-c{\cal T}(x_l))
	\cos(\varphi-\alpha(x_l))
    =0.
\end{eqnarray}

\noindent We have used (\ref{taul}) to eliminate $\tau_l$ in terms of $x_l$ and
$y_l$.

The envelope is obtained by setting $\partial G(x,y,t,\varphi)/\partial\varphi
=0$, where the partial derivative is taken at constant values of $(x,y,t)$.
On the observer side, i.e., for $y<y_l$, taking $\partial/\partial\varphi$
on (\ref{xala}) we obtain

\begin{equation}
	-y\sin\varphi +x\cos\varphi = 0
\end{equation}

\noindent or

\begin{equation}\label{varphixa}
	\tan\varphi = \frac{x}{y}, 
\end{equation}

\noindent an equation for $\varphi$ as a function of $x^a$ on the observer's
side. Substituting (\ref{varphixa}) back into (\ref{xala}) we
have (for $y<y_l$),

\begin{equation}  \label{lc}
	\sqrt{x^2+y^2}+c(t-T_0) =0
\end{equation}

\noindent which is the observer's past lightcone, as expected. Notice that
when (\ref{varphixa}) is evaluated at the lens plane it yields 

\begin{equation}\label{varphixaimplicit}
	\tan\varphi =\frac{x_l}{y_l}.  
\end{equation}

\noindent From Eq.~(\ref{s}), we see that this implies that $r=0$, i.e., that
these rays come from the observer. In other words, of all the lightrays at
angle $\varphi$, the envelope condition picks the null ray that passes
through the observer. This lightray hits the lens plane at $x=x_l$ and
subsequently bends through an as yet undetermined angle $\alpha(x_l)$. In
order to determine the bending angle $\alpha$ we use the envelope condition
on the source's side, imposing on it the condition (\ref{varphixaimplicit}).

The envelope is obtained by setting to zero the implicit differentiation of
$G(x,y,\tau,\varphi)$ from Eq.~(\ref{xalasource}) with respect to $\varphi$
keeping $x^a$ fixed. We obtain:

\begin{eqnarray}
	0 
	&=&   \sin(\varphi-\alpha)
	      \left[\frac{\partial x_l}{\partial \varphi}
	            \frac{\partial\alpha}{\partial x_l}
                    -1\right]
	     (cT_0-ct-x_l\sin\varphi-y_l\cos\varphi-c{\cal T})  \nonumber \\
	& & +\cos(\varphi-\alpha)
	     \left[ y_l\sin\varphi
                   -x_l\cos\varphi
	           -\frac{\partial x_l}{\partial\varphi}
	           \left(c\frac{\partial{\cal T}}{\partial x_l}
		         +\sin\varphi\right)
	     \right].					\label{envelopeprelim}
\end{eqnarray}

The quantity $\frac{\partial x_l}{\partial\varphi}$ is calculated by
implicit differentiation of (\ref{fermatsx}) with respect to $\varphi$,
keeping $x^a$ fixed:

\begin{equation}\label{A}
\frac{\partial x_l}{\partial\varphi}= -\frac{A}{B}, 
\end{equation}

\noindent with
\begin{mathletters}
\begin{eqnarray}
    A &=& \cos(\varphi-\alpha)
	      (cT_0-ct-x_l\sin\varphi-y_l\cos\varphi-c{\cal T}) \nonumber\\
      & &+\sin(\varphi-\alpha)
	      (y_l\sin\varphi-x_l\cos\varphi)\\
    B &=&1
	 -\cos(\varphi-\alpha)
	  (cT_0-ct-x_l\sin\varphi-y_l\cos\varphi-c{\cal T})
	  \frac{\partial\alpha}{\partial x_l} \nonumber\\
      & &-\sin(\varphi-\alpha)
	  \left(c\frac{\partial{\cal T}}{\partial x_l}
                +\sin\varphi \right).  
\end{eqnarray}
\end{mathletters}

\noindent From Eq.(\ref{A}), and some algebra, Eq.~(\ref{envelopeprelim})
becomes

\begin{eqnarray}
   0 &=&-\sin(\varphi-\alpha)
	+c\frac{\partial{\cal T}}{\partial x_l}
	+\sin\varphi  						\nonumber\\
     & &+(y_l\sin\varphi-x_l\cos\varphi)
	 \left[\frac{\cos(\varphi-\alpha)}
	            {(cT_0-ct-x_l\sin\varphi-y_l\cos\varphi-c{\cal T})}
	       -\frac{\partial\alpha}{\partial x_l}\right] .  \label{envelope}
\end{eqnarray}

\noindent Using Eq.~(\ref{varphixaimplicit}), the first parenthesis in the last
term of Eq.~(\ref{envelope}) vanishes, so that Eq.~(\ref{envelope}) reduces to

\begin{equation}\label{timeanglelarge}
   0 =   \sin\varphi 
	-\sin(\varphi-\alpha)
	+c\frac{\partial{\cal T}}{\partial x_l}.  
\end{equation}

\noindent This is the basic relationship between the bending angle and the
gravitational time delay that holds for large as well as small angles. Using
the regime of small angles, Eq.~(\ref{timeanglelarge}) reduces to

\begin{equation}\label{timeangle}
   \alpha =-c\frac{\partial{\cal T}}{\partial x_l},  
\end{equation}

\noindent the standard relationship in the astrophysical approach to lensing
in a two-dimensional scenario.

The standard lens equation is obtained in the following manner. Elimination of
$\tau-\tau_l-{\cal T}$ in Eqs.~(\ref{fermats}) implies

\begin{equation}
(x-x_l)\cos(\varphi-\alpha)-(y-y_l)\sin(\varphi-\alpha) = 0.
\end{equation}

\noindent With small angles, and with $y_l=D_l$ and $y=D_s$ (the source
distance), this becomes

\begin{equation}
   x-x_l-(D_s-D_l)(\varphi-\alpha) = 0.
\end{equation}

\noindent From the envelope condition, Eq.~(\ref{varphixaimplicit}), with the
small-angle approximation ($\varphi =\tan\varphi=x_l/D_l$), we
finally have

\begin{equation}  \label{stand-thin-lens}
  \frac{x}{D_s} = \frac{x_l}{D_l} - \frac{(D_s-D_l)}{D_s}\alpha,
\end{equation}

\noindent the standard lens equation, which is understood as a map from $x_l$
on the lens line to $x$ on the source line (i.e., $x_l\to x$), with fixed
values of $D_l$ and $D_s$.

\begin{figure}
\centerline{\hbox{\vbox{\psfig{figure=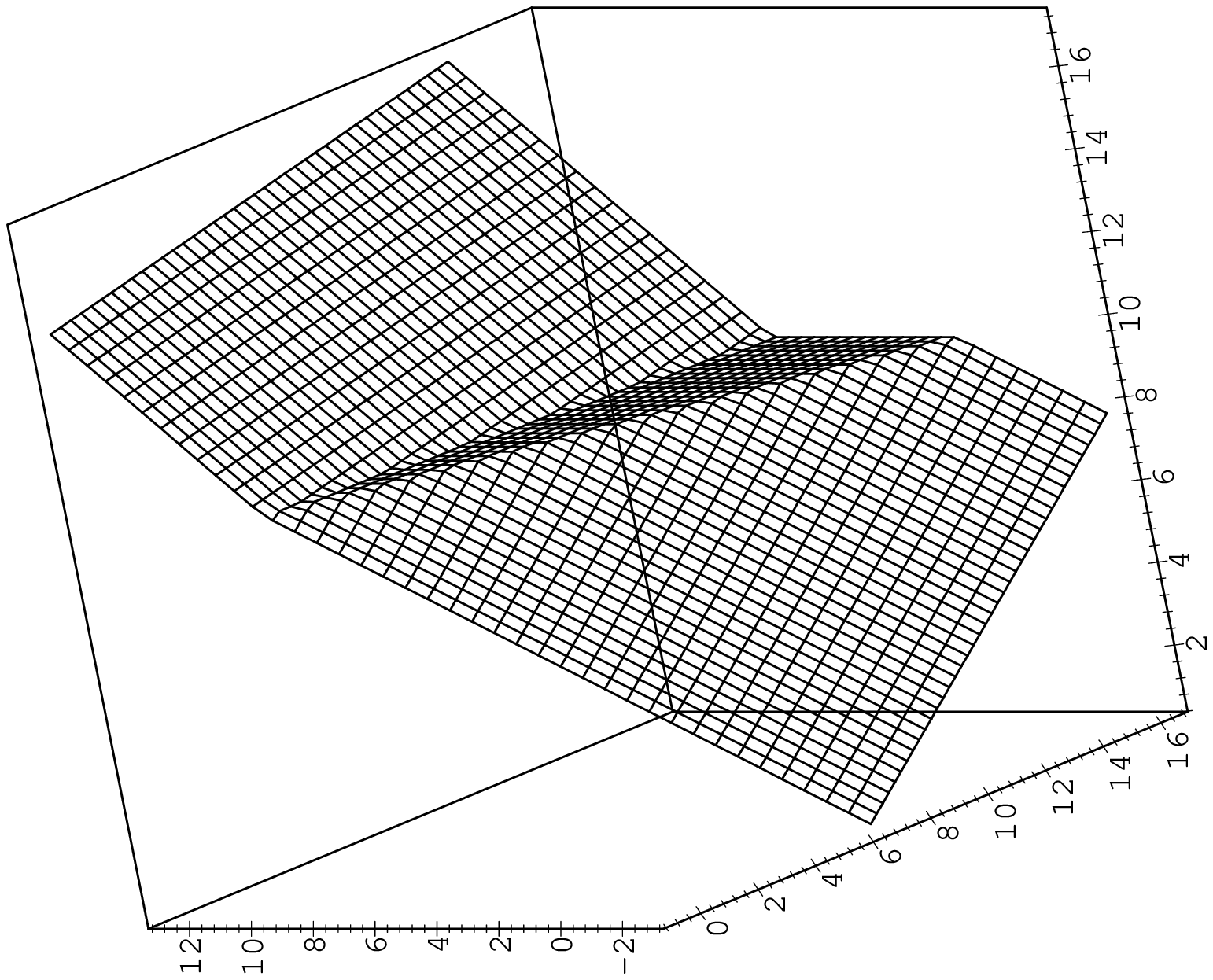,width=3in,angle=-90}
          \psfig{figure=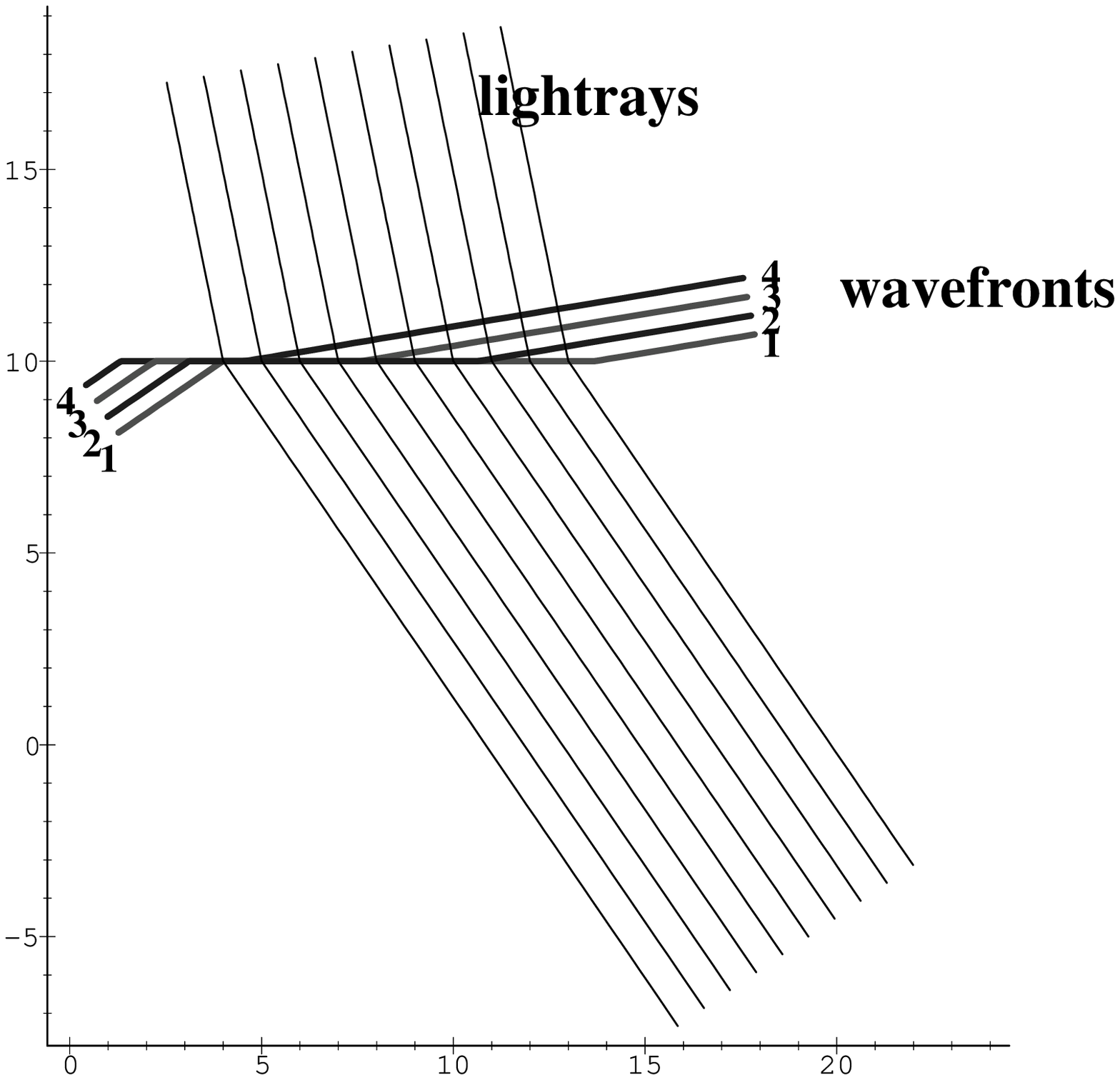,width=3in,angle=-90}
          \psfig{figure=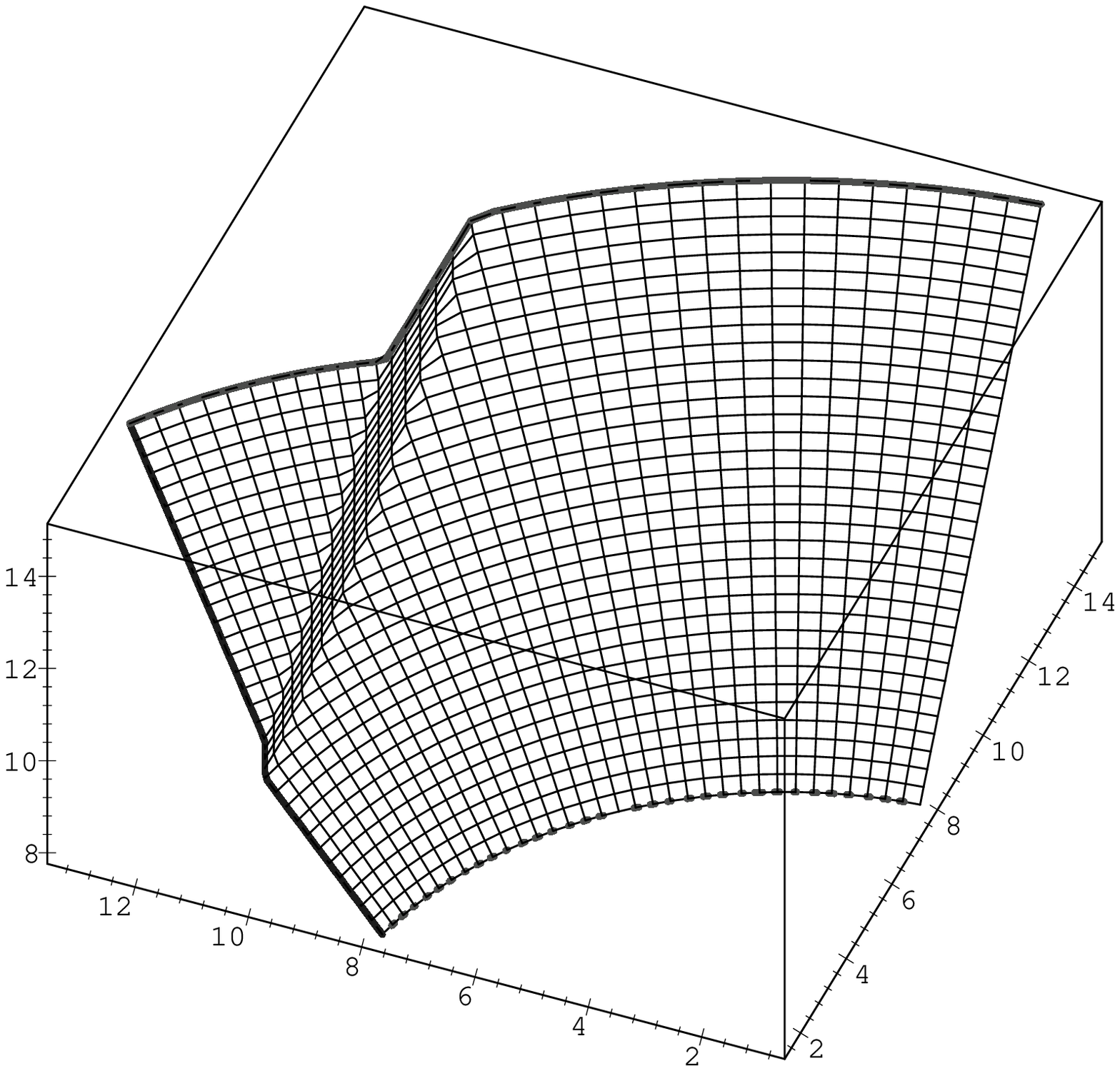,width=3in,height=3in,angle=-90}
               }
        }
}

\caption{Envelope construction for the case of a bending angle
independent of the location on the lens line $x_l$.  The top panel shows
a spacetime picture of one member of the 1-parameter family of null
surfaces that is used to construct the envelope. The center panel is a
space picture of the null surface, where both the lightrays and the
associated wavefronts are shown. The bottom panel shows a spacetime
picture of the envelope of the 1-parameter family. The family is labeled
by the direction of the plane wave that hits the lens line.  }

\label{fig:1}
\end{figure}

As an illustration, Figure\ref{fig:1} shows how the envelope construction
works for the case of a bending angle that does not depend on the point on
the lens plane: $\alpha(x_l)=const.$ and $c{\cal T}=-\alpha x_l$.
The top panel shows a spacetime picture of one member of the 1-parameter
family of null surfaces that is used to construct the envelope. In this
case, two null planes in the directions $\varphi$ and $\varphi-\alpha $
are matched at the lens' worldsheet. The vertical separation between the
intersections of the two null planes with the lens' worldsheet represents
the time delay ${\cal T}$. A spatial projection of the resulting null
surface shows the bending of lightrays, whereas the time delay is observed
in the shift of the wavefronts in passing through the lens line. The
envelope of the $\varphi -$dependent family of such null surfaces yields the
lightcone of the observer on the observer's side of the lens, which is
matched to another lightcone with a shifted origin on the other side of the
lens.

\begin{figure}
\centerline{\hbox{\vbox{\psfig{figure=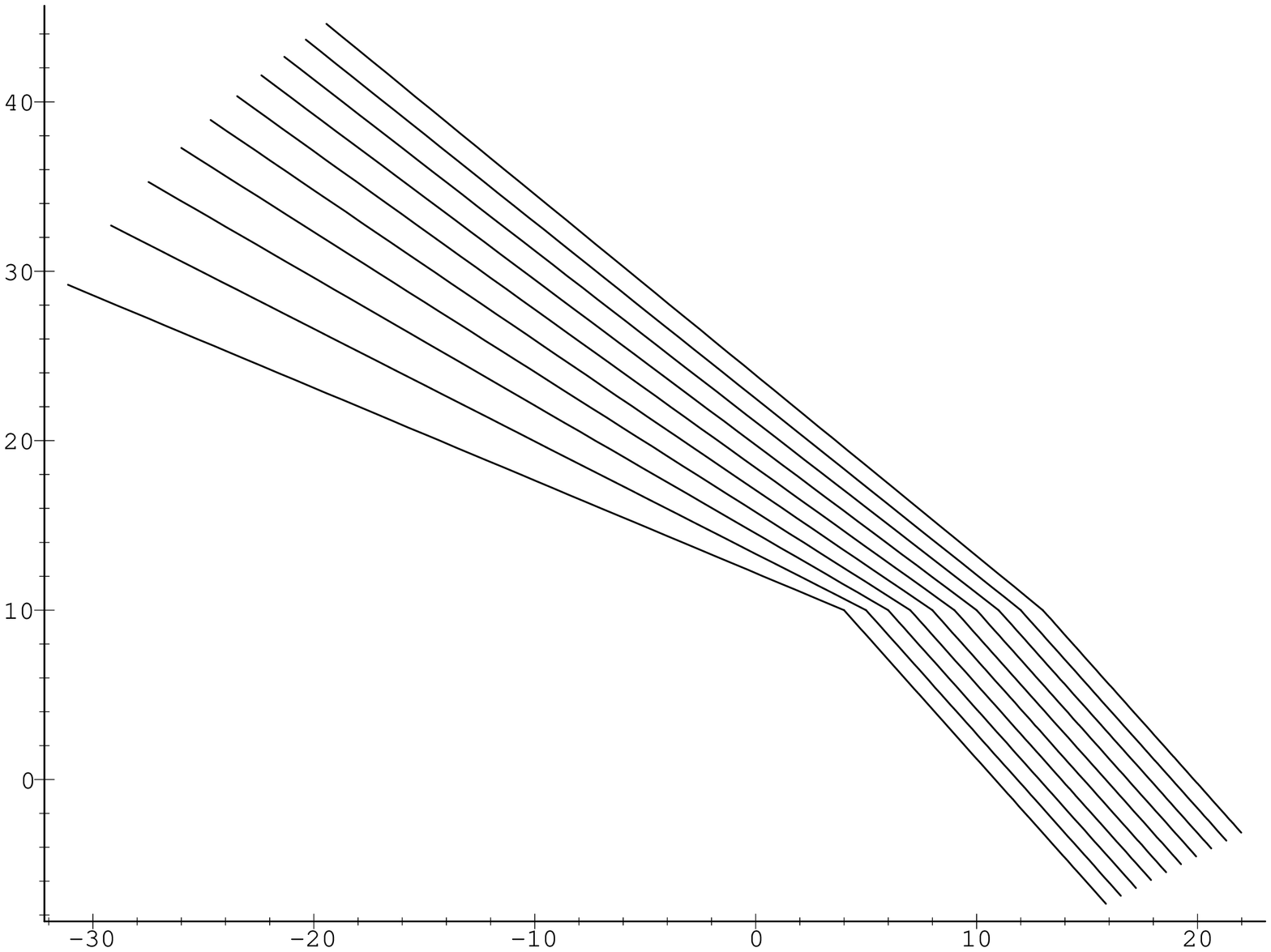,width=3in,angle=-90}
          \psfig{figure=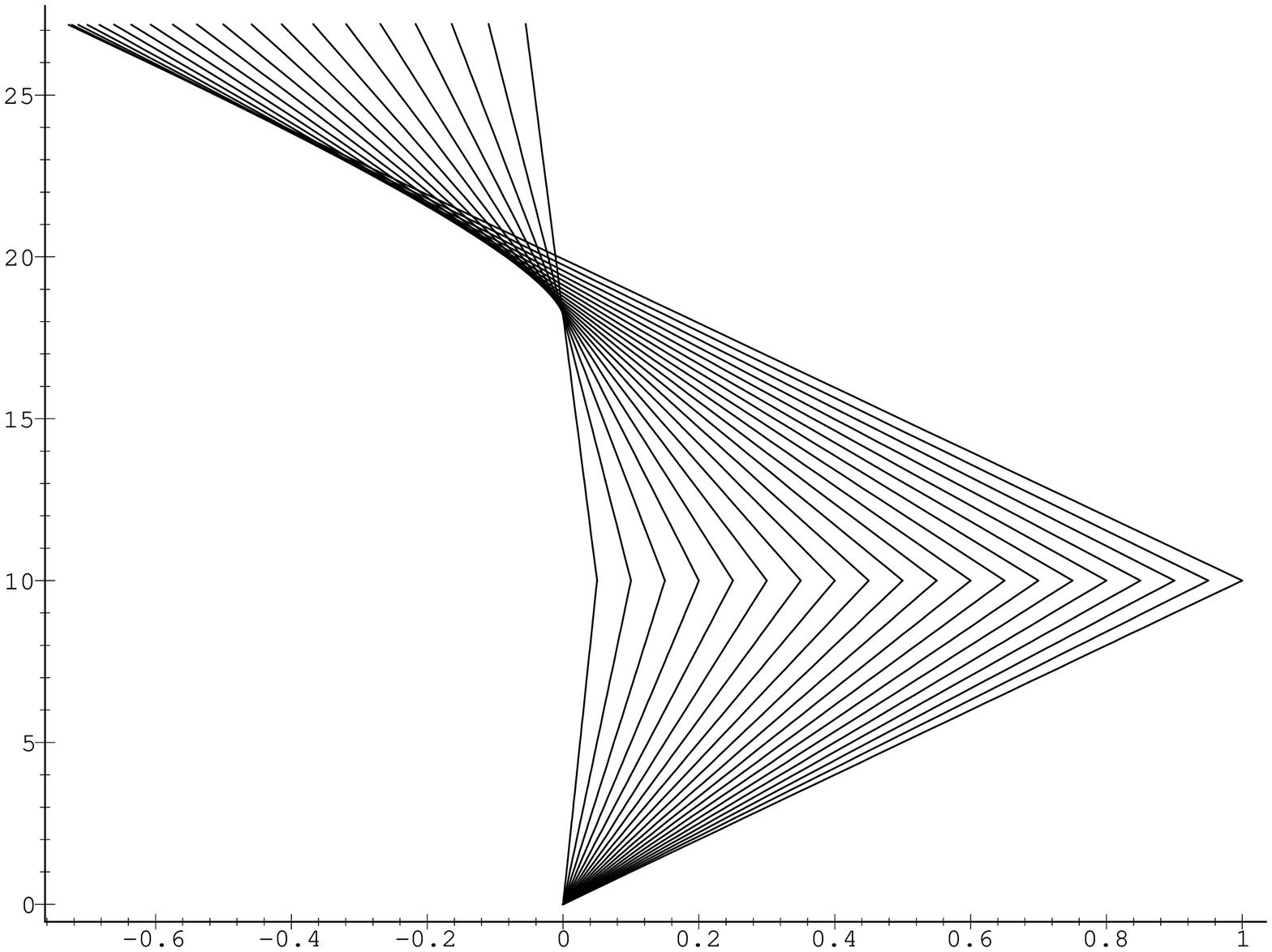,width=3in,angle=-90}
          \psfig{figure=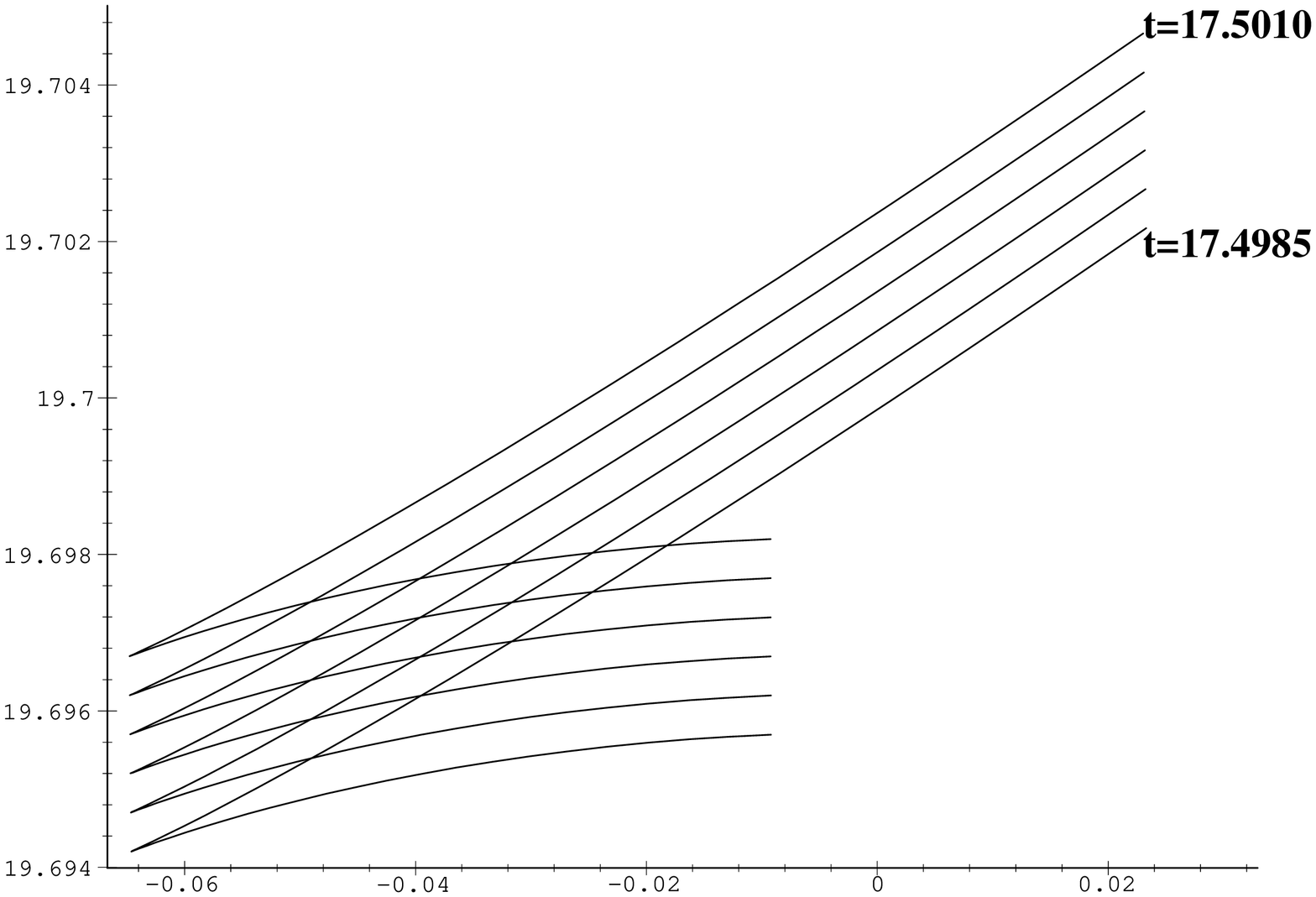,width=3in,angle=-90}}
        }
}

\caption{Envelope construction for the case of a time delay
$T(x_l)=T_0-\ln(1+(x_l/x_0)^2)$.  The top panel shows a space projection of one
member of the family of null surfaces. The null surfaces are labeled by the
direction of the plane that hits the lens line. The center panel contains a
space picture of the envelope showing the lightrays. Notice the caustic line
that develops after the lightrays pass through the lens line.  The bottom panel
shows a greatly magnified view of a few of the wavefronts associated with the
lightrays of the center panel. Notice the cusp in each wavefront. These
wavefronts occur around the value 20 in the verical axis of the center panel. }

\label{fig:2}
\end{figure}

A more physical example is provided by the use of a time delay of the form
${\cal T}(x_l)={\cal T}_0(1-\ln (1+(x_l/x_0)^2))$. This reproduces
qualitatively the time delay of a compact lens in the vicinity of the optical
axis, since the delay is greater for smaller impact parameters. The spatial
projection of the null surface obtained in this case is shown in the top panel
of Fig. \ref{fig:2}. The center panel of the figure shows the projection of the
envelope. In this case, a caustic develops on the source's side of the
observer's lightcone. The caustic is traced by cusps in the wavefronts, as
shown in the bottom panel of the figure.

The equation for the time delay $\Delta\tau$ (the difference in arrival time
between a lensed light ray and an unlensed one) can be obtained from our
approach by solving for $\tau$ in (\ref{fermatsy}) and subtracting from it the
travel time in the absence of the lens:

\begin{equation}\label{timedelay1}
   \Delta\tau =  \frac{y-y_l}{c\cos(\varphi-\alpha)}
		+\tau_l+{\cal T}
		-\frac{1}{c}\sqrt{x^2+y^2}
\end{equation}

\noindent The equation for the time delay is usually given up to second order
in the angles. We need expansions of $x_l$ and $\tau_l$, from
Eqs.~(\ref{fermato}) with $r=0$, up to second-order
terms:

\begin{mathletters}
\begin{eqnarray}
   \tau_l &=& \frac{y_l}{c}\left[1+\frac{\varphi^2}{2}\right] 
	     + O(\varphi^3),\\
      x_l &=& y_l\varphi +O(\varphi^3).
\end{eqnarray}
\end{mathletters}

\noindent These, as well as (\ref{stand-thin-lens}) for $x$, are used in
(\ref{timedelay1}). We obtain

\begin{equation}\label{usualdeltatau}
   \Delta\tau ={\cal T}+\frac{D_lD_{ls}}{2cD_s}\alpha^2,
\end{equation}

\noindent which agrees with the usual expression for the time delay.

\subsection{3+1 static lens planes}

The generalization of our scheme for the thin-lens spacetimes to 3+1 dimensions
is straightforward. We consider a spacetime with coordinates $(t,x,y,z)$
consisting of two Minkowski spacetimes matched together at the lens plane,
which in this case is the timelike 3-surface $z=z_l$ and is parametrized by
$(t,x_l,y_l)$. A generic null direction is $l^a$ given in terms of
stereographic coordinates as in Eq.~(\ref{ell}). For our present purposes, we
relabel the components of $l^a$ as

\begin{equation}\label{p}
    	l^a = \frac{1}{\sqrt{2}}(1,p_x,p_y,p_z)  
\end{equation}

\noindent where $(p_x,p_y)$ label the sphere of null directions and 
$p_z \equiv \sqrt{1-p_x^2-p_y^2}$. As in the (2+1)--case, we choose
our family of null surfaces as null planes on the observer's side of the
spacetime, and write it parametrically as

\begin{equation}\label{3+1.1}
    \vec{r} = c\tau\vec{p}+\vec{s} 
\end{equation}

\noindent with $\vec{r}\equiv (x,y,z)$, $\vec{p}\equiv (p_x,p_y,p_z)$
and $\vec{s}=(s_x,s_y,s_z)$, with $\vec{s}\cdot \vec{p}=0$,
representing the null surface at $\tau =0$. The values of $\vec{s}$ label
the null geodesics of the $\vec{p}$ null plane.

Then, if $\tau_l$ represents the time that it takes for a lightray to
reach the point $\vec{r}_l\equiv (x_l,y_l,z_l)$ measured from the
time that the wavefront passes the observer, we must have

\begin{equation}\label{rl}
   \vec{r}_l = c\tau_l\vec{p}+\vec{s}  
\end{equation}

\noindent and thus

\begin{equation}\label{taul-rl}
   \tau_l = \frac{\vec{r}_l\cdot\vec{p}}{c}. 
\end{equation}

\noindent It is very useful to change the geodesic labels from $\vec{s}$ to
$\vec{r}_l$ by means of 

\begin{equation}\label{s-rl}
   \vec{s} = \vec{r}_l-(\vec{r}_l\cdot\vec{p})\vec{p}.  
\end{equation}

\noindent for use on both the observer and source side of the spacetime.
Eq.~(\ref{3+1.1}) becomes 

\begin{equation}\label{3+1.0}
   \vec{r} = \big(c\tau -(\vec{r}_l\cdot\vec{p})\big)\vec{p}+\vec{r}_l.
\end{equation}

\noindent We thus have in this picture that the wavefront, in direction
$\vec{p}$, consists of null geodesics parallel to each other, parametrized by
the values of $(x_l,y_l)$ where they hit the lens plane at $z_l$. Once a ray
reaches the lens plane, it is delayed for a time ${\cal T}(x_l,y_l)$, during
which it just ``sits'' at the lens plane, before leaving in a different
direction $\vec{p}'$. Thus, in our model, we have

\begin{equation}\label{3+1.2}
   \vec{r} = \vec{r}_l  
\end{equation}

\noindent and then

\begin{equation}\label{3+1.3}
   \vec{r} = c(T_0-t-\tau_l-{\cal T})\vec{p}'+\vec{r}_l
\end{equation}

\noindent where

\begin{equation}
   \vec{p}' = \vec{p}-\vec{\alpha}(x_l,y_l).  \label{p'}
\end{equation}

\noindent Here we have defined $\vec{\alpha}(x_l,y_l)$ in the usual manner,
representing the deviation of the null geodesic, which depends on the point at
which the geodesic hits the lens plane. Equations (\ref{3+1.0}), (\ref{3+1.2}),
and (\ref{3+1.3}) constitute a parametric expression of 
$G(t,x,y,z,p_x,p_y)=0$. In this approach, the time delay  ${\cal T}(x_l,y_l)$
is assumed to be prescribed, whereas $\vec{\alpha}(x_l,y_l)$ is determined by
the envelope condition. The envelope condition in this case takes the form

\begin{equation}
\frac{\partial G}{\partial p_x} =\frac{\partial G}{\partial p_y} =0
\end{equation}

\noindent which is carried out implicitly as in the (2+1)-case, and which
yields

\begin{mathletters}
\begin{eqnarray}
	c\frac{\partial{\cal T}}{\partial x_l} 
    &=& -\alpha_x(x_l,y_l), \\
	c\frac{\partial{\cal T}}{\partial y_l} 
    &=& -\alpha_y(x_l,y_l),
\end{eqnarray}
\end{mathletters}

\noindent thus reproducing the relationship between the time delay and the
bending angle in the more standard astrophysical approach to lensing. The
detailed calculations, which can be omitted, are reproduced below.

The implicit $G$ function on the observer's side is given by the $z-$component
of Eq.~(\ref{3+1.0})

\begin{equation}\label{G-before1}
	G(\tau,x,y,z,p_x,p_y) = z-z_l-(\tau-\tau_l)p_z = 0,
\end{equation}

\noindent while $x_l$ and $y_l$ are defined implicitly by the
$(x,y)-$components of (\ref{3+1.0}). Solving for $x_l$ and $y_l$ from these
equations and substituting back into $G$ yields

\begin{equation}
   G = \frac{1}{p_z}(-t+xp_x+yp_y+zp_z) = 0. 
\end{equation}

\noindent On the source side we implicitly define the thin-lens $G$ function
by

\begin{equation}\label{G-afterL}
   G = z-z_l-(\tau -\tau_l-{\cal T})p_z' = 0,  
\end{equation}

\noindent where $x_l$ and $y_l$ are implicitly defined by the
$(x,y)-$components of Eq.(\ref{3+1.3}). The envelope of $G$ is obtained by
setting to zero the partial derivatives of $G$ with respect to $p_x$ and
$p_y$, holding $(t,x,y,z)$ fixed. Applying this prescription to the $G$
function before the lens plane, we have

\begin{mathletters}\label{3+1plane-p}
\begin{eqnarray}
   \frac{\partial G}{\partial p_x} = 0 
 \quad &\Rightarrow& \quad 
    x - \frac{p_x}{p_z} z = 0,  		\label{env-before-x} \\
   \frac{\partial G}{\partial p_y} = 0 
 \quad &\Rightarrow& \quad 
    y - \frac{p_y}{p_z} z = 0. 			\label{env-before-y}
\end{eqnarray}
\end{mathletters}

\noindent We note that Eqs.~(\ref{3+1plane-p}) imply that

\begin{equation}  \label{env-before}
   \frac{x_l}{p_x} = \frac{y_l}{p_y} = \frac{z_l}{p_z}.
\end{equation}

\noindent Solving for $p_x$ and $p_y$ from Eqs.~(\ref{3+1plane-p}) and
substituting into the $G$ function, we obtain our final expression for the
thin-lens $G$ function on the observer's side

\begin{eqnarray}
   p_z G &=&-t+xp_x+yp_y+zp_z
	   =-t+\frac{p_z}{z}(x^2+y^2+z^2) \nonumber  \\
         &=&-t\pm \sqrt{x^2+y^2+z^2}=0,  \label{G-before-fin}
\end{eqnarray}

\noindent the observer's lightcone.

On the source's side, the envelope construction requires computing the
following partial derivatives: $\frac{\partial x_l}{\partial p_x},
\frac{\partial y_l}{\partial p_x}, \frac{\partial x_l}{\partial p_y}$ and 
$\frac{\partial y_l}{\partial p_y}$ from the $(x,y)-$components of
Eq.~(\ref{3+1.3}). Since, $\vec{p}'=\vec{p}-\vec{\alpha}(x_l,y_l)$, we need the
derivatives of $\vec{p}'$.

\noindent We define

\begin{equation}  \label{dp}
   	  \frac{\delta p'_x}{\delta p_x} 
   \equiv \frac{\partial p'_x}{\partial p_x} 
	 +\frac{\partial p'_x}{\partial x_l} 
	  \frac{\partial x_l}{\partial p_x} 
	 +\frac{\partial p'_x}{\partial y_l} 
	  \frac{\partial y_l}{\partial p_x},
\end{equation}

\noindent and similar quantities for all combinations of $(p'_x,p'_y,p_x,p_y)$.
After a lengthy calculation, we obtain

\begin{mathletters}
\begin{eqnarray}
	\frac{\partial x_l}{\partial p_x} 
    &=& \frac{(\tau -\tau_l-{\cal T})
	   \left\{-\frac{\delta p'_x}{\delta p_x}
	   	\left(1-\frac{\partial{\cal T}}{\partial y_l}p'_y
		-p'_y p_y\right) 
		  -\frac{\delta p'_y}{\delta p_x}
		\left(\frac{\partial{\cal T}}{\partial y_l}p'_x
		+p'_xp_y\right) \right\} }
	     {1-p'_x\frac{\partial{\cal T}}{\partial x_l}
	       -p'_y\frac{\partial{\cal T}}{\partial y_l}
	       -p_xp'_x-p_yp'_y},  			\label{xpx2} \\
	\frac{\partial y_l}{\partial p_x} 
   &=&  \frac{(\tau -\tau_l-{\cal T})
	   \left\{-\frac{\delta p'_y}{\delta p_x}
		\left(1-\frac{\partial{\cal T}}{\partial x_l}p'_x
		-p'_xp_x\right) 
		  -\frac{\delta p'_x}{\delta p_x}
		\left(\frac{\partial{\cal T}}{\partial x_l}p'_y
		+p'_yp_x\right) \right\} }
	     {1-p'_x\frac{\partial{\cal T}}{\partial x_l}
	       -p'_y\frac{\partial{\cal T}}{\partial y_l}
	       -p_xp'_x-p_yp'_y}.  \label{ypx2}
\end{eqnarray}
\end{mathletters}

\noindent and similar expressions for $\partial x_l/\partial p_y$ and  $\partial
y_l/\partial p_y$.

The envelope construction consists in setting the $p_x$ and $p_y$ partial
derivatives (holding $(t,x,y,z)$ fixed) of Eq.~(\ref{G-afterL}) to zero.
Explicitly, we have

\begin{mathletters}
\begin{eqnarray}
0 &=&-(\tau -\tau_l-{\cal T})\left( \frac{\partial p_{z}^{\prime }}{
\partial p_{x}^{\prime }}\frac{\delta p_{x}^{\prime }}{\delta p_{x}}+\frac{
\partial p_{z}^{\prime }}{\partial p_{y}^{\prime }}\frac{\delta
p_{y}^{\prime }}{\delta p_{x}}\right)  \nonumber \\
&&+p_{z}^{\prime }\left( p_{x}\frac{\partial x_l}{\partial p_{x}}+p_{y}
\frac{\partial y_l}{\partial p_{x}}+\frac{\partial {\cal T}}{\partial
x_l}\frac{\partial x_l}{\partial p_{x}}+\frac{\partial {\cal T}}{
\partial y_l}\frac{\partial y_l}{\partial p_{x}}\right) ,  \\
0 &=&-(\tau -\tau_l-{\cal T})\left( \frac{\partial p_{z}^{\prime }}{
\partial p_{x}^{\prime }}\frac{\delta p_{x}^{\prime }}{\delta p_{y}}+\frac{
\partial p_{z}^{\prime }}{\partial p_{y}^{\prime }}\frac{\delta
p_{y}^{\prime }}{\delta p_{y}}\right)  \nonumber \\
&&+p_{z}^{\prime }\left( p_{x}\frac{\partial x_l}{\partial p_{y}}+p_{y}
\frac{\partial y_l}{\partial p_{y}}+\frac{\partial {\cal T}}{\partial
x_l}\frac{\partial x_l}{\partial p_{y}}+\frac{\partial {\cal T}}{
\partial y_l}\frac{\partial y_l}{\partial p_{y}}\right) , 
\end{eqnarray}
\end{mathletters}

\noindent where we have again used Eq.~(\ref{env-before}). In these two
equations, we substitute the partial derivatives given above and solve for
$\partial{\cal T}/\partial x_l$ and $\partial{\cal T}/\partial y_l$, obtaining

\begin{mathletters}\label{gradT}
\begin{eqnarray}
   	\frac{\partial{\cal T}}{\partial x_l} 
   &=&	p'_x-p_x = \alpha_x(x_l,y_l),  		\label{dTdx} \\
	\frac{\partial{\cal T}}{\partial y_l} 
   &=&  p'_y-p_y=\alpha_y(x_l,y_l).  		\label{dTdy}
\end{eqnarray}
\end{mathletters}

Eqs.~(\ref{gradT}) determine the new direction of the light ray on the source's
side with respect to the initial direction of the light ray on the observer's
side (i.e., the bending angle) as the gradient of the time delay function. Note
that we have not made use of the small angle approximation in these calculations.

\section{Moving lens planes in 2+1 dimensions}


As an application of our approach, we consider a generalization of our
(2+1)--results to the case of a lens plane that is moving in the
observer's frame of reference, but is otherwise unchanging. In other words, the
deflector has a given configuration, which does not change with time, but moves
rigidly along the line of sight. The idea is to obtain the bending angle and
compare with the bending angle by the same deflector at rest.

Again our basic picture is a family of parallel null geodesics at an angle
$\varphi$ with the normal to the lens plane forming a null plane. The rays each
arrive at the lens plane at a time $\tau_l$, and are then remain in the lens
plane for a length of time ${\cal T}$ that depends only on the point at which
they hit the now moving lens plane, before exiting the lens plane at an
angle $\varphi'=\varphi-\alpha_m$.

We point out that the ``detention time'' ${\cal T}$ of a thin rigid lens in
(slow) motion is the same as that of the same deflector at rest. The reason for
this becomes clear if we consider the situation of a glass sheet of thickness
$\Delta$ that is moving. In the rest frame of the glass, light has a speed $u_0$
and takes a time $t_0=\Delta /u_0$ between entering and exiting the glass. The
time in the laboratory frame in which the glass is moving with speed $v$ is
$t'=t_0-v\Delta/c^2$ (neglecting higher powers in $v/c$) following from the fact
that the length for light to travel changes due to the motion of the glass, and
with the fact that the speed of light also changes if the glass is moving (by the
Fizeau effect). Thus $t'$ differs from $t_0$ only by a term proportional to the
thickness of the glass, but completely unrelated to the glass properties. In the
limit in which the thickness of the glass tends to zero while the rest-frame
travel time, $t_0$ is kept constant, the difference vanishes.

As before, we assume that the observer lies at the origin of coordinates 
$x_0^a=(T_0,0,0)$. The lens plane, however, is moving, being described
by $x_l^a=(t,x_l,y_l(t))$. The (affine) parameter $\tau$
describing the evolution of the null geodesics of the null surface is such
that the wavefront $\tau=0$ passes by the observer at $t=T_0$. The
parameter $\tau$ runs backwards in time, so that the time coordinate of the
lightrays is given by $t=T_0-\tau$ as before. We will slightly abuse our
notation by using $y_l(\tau)$ to denote $y_l(t=T_0-\tau)$. If the
individual lightrays in the beam are labeled by $r$, with $r=0$ being the
ray that passes through the origin, then we can give them parametrically as

\begin{mathletters}\label{vobs}
\begin{eqnarray}
	x &=& r\cos\varphi +c\tau \sin\varphi ,  \label{xobs}\\
	y &=&-r\sin\varphi +c\tau \cos\varphi .  \label{yobs}
\end{eqnarray}
\end{mathletters}

\noindent Individual lightrays reach the lens plane at time $\tau_l$, at
which

\begin{mathletters}\label{lensintersec}
\begin{eqnarray}
    x_l         &=& r\cos\varphi +c\tau_l\sin\varphi  \label{lensintersec1}\\
    y_l(\tau_l) &=&-r\sin\varphi +c\tau_l\cos \varphi \label{lensintersec2}
\end{eqnarray}
\end{mathletters}

\noindent The rays remain at the lens plane for a time ${\cal T}(x_l)$ and then
leave in a direction $\varphi-\alpha_m(x_l)$, where the bending angle for a
moving lens, $\alpha_m(x_l)$, is to be determined. At the time the rays leave the
lens plane, however, the lens plane has moved to a point $y_l(\tau_l+{\cal T})$,
carrying the rays with it. On the other side of the lens, the source side, we
have

\begin{mathletters}\label{vsource}
\begin{eqnarray}
   x &=& c(\tau -\tau_l-{\cal T})\sin(\varphi-\alpha_m)+x_l,  \label{xsource}\\
   y &=& c(\tau -\tau_l-{\cal T})\cos(\varphi-\alpha_m)+y_l(\tau_l+{\cal T}).  
							      \label{ysource}
\end{eqnarray}
\end{mathletters}

\noindent Eqs.~(\ref{vobs}) hold when $\tau <\tau_l$, whereas
Eqs.~(\ref{vsource}) hold for $\tau >\tau_l+{\cal T}$. On the observer side
(\ref{yobs}) represents $G(x,y,t,\varphi)=0$ if $r$ is thought of as a function
of $(t,x,\varphi)$ given implicitly by (\ref{xobs}). The envelope condition
$\partial G/\partial\varphi =0$ on the observer side, implemented by taking
$\partial/\partial\varphi$ of (\ref{xobs}), yields, as we had earlier, $r=0$
or

\begin{mathletters}\label{r=0}
\begin{eqnarray}
	x_l         &=&c\tau_l\sin\varphi ,  \label{r=01} \\
	y_l(\tau_l) &=&c\tau_l\cos\varphi .  \label{r=02}
\end{eqnarray}
\end{mathletters}

\noindent As earlier, we consider (\ref{ysource}) as the equation
$G(x,y,t,\varphi)=0$ on the source's side by interpreting $x_l$ as a function of
$(\tau,x,\varphi)$:

\begin{equation}
	x_l  = x_l(\tau,x,\varphi),
\end{equation}

\noindent given implicitly by (\ref{xsource}). The implementation of the
envelope condition, $\partial G/\partial\varphi=0,$ involves several steps that
must be explained.

First, we have to evaluate $\frac{dy_l}{d\tau}$, the velocity of the lens
when the light first enters it, at $\tau_l+{\cal T}$ and when it
exits the lens, at $\tau_l$, on the observer side. Since the lens is thin
and moving slowly, we consider the two values to be the same (or equal to an
average) and call

\begin{equation}\label{vl}
   \frac{dy_l}{d\tau} = -\widehat{v}_l.  
\end{equation}

The second point is that, though Eq.~(\ref{ysource}) contains $x_l$ via 
$\alpha$ and ${\cal T}$, it also contains $\tau_l$ which is a function
of both $x_l$ and $\varphi$ and is thus needed in the envelope condition.
It is obtained by eliminating $r$ from Eqs.~(\ref{vobs}):

\begin{equation}\label{taulv}
	c\tau_l-x_l\sin\varphi-y_l(\tau_l)=0.  
\end{equation}

\noindent Taking the $\varphi$ derivative of Eq.~(\ref{taulv}) and using
(\ref{vl}), with $r=0$, yields

\begin{equation}\label{taulphi}
   \frac{\partial\tau_l}{\partial\varphi}
  =\frac{\partial x_l}{\partial\varphi}
   \frac{\sin\varphi}{(c+\cos\varphi\widehat{v}_l)}.
\end{equation}

Next we need the $\varphi-$derivative of $x_l$, calculated from (\ref{xsource}),
which, using (\ref{taulphi}), yields

\begin{equation}\label{dxldphi}
	\frac{\partial x_l}{\partial\varphi}
      = \frac{ -c(\tau -\tau_l-{\cal T})\cos(\varphi-\alpha_m)}
	     {1-c(\tau -\tau_l-{\cal T})\cos(\varphi-\alpha_m)}
	\frac{\partial\alpha_m}{\partial x_l}
	-c\left(\frac{\partial{\cal T}}{\partial x_l}
	       +\frac{\sin\varphi}{(c+\widehat{v}_l\cos\varphi)}
	  \right)\sin(\varphi -\alpha_m).  
\end{equation}

Finally, setting $\partial G/\partial\varphi =0$ from (\ref{ysource})
yields

\begin{eqnarray}
0 &=&\left(\frac{\partial\tau_l}{\partial\varphi}
	  +\frac{\partial {\cal T}}{\partial x_l}
	   \frac{\partial x_l}{\partial\varphi}\right)
     \left(c\cos(\varphi-\alpha_m)+\widehat{v}_l\right)  \nonumber\\
  & &+c(\tau -\tau_l-{\cal T})\sin(\varphi-\alpha_m)
	\left(1-\frac{\partial\alpha_m}{\partial x_l}
		\frac{\partial x_l}{\partial\varphi}
	\right) 					\label{dGdphi}			
\end{eqnarray}

which, using (\ref{taulphi}), (\ref{dxldphi}) and the fact that $x_l\cos\varphi
-\sin\varphi y_l(\tau_l)=0$ (from Eqs.(\ref{r=0})) becomes,

\begin{equation}\label{prelimalpha}
 \sin(\varphi-\alpha_m)
-\left(c+\widehat{v}_l\cos(\varphi-\alpha_m)\right) 
 \left(\frac{\partial{\cal T}}{\partial x_l}
      +\frac{\sin\varphi}{(c+\widehat{v}_l\cos\varphi)}\right) =0.  
\end{equation}

Now, using the results of the envelope condition on the observer side where
$\varphi$ is a function of $x_l$ given implicitly by eliminating  $\tau_l$ from
Eqs.~(\ref{r=0}), we can interpret Eq.(\ref{prelimalpha}) as an equation defining
the bending angle, $\alpha_m$, as a function of $x_l$ if ${\cal T}(x_l)$ is
known. In the regime of small angles, keeping only linear terms in $\varphi $ and
$\alpha_m$, it reduces to

\begin{equation}\label{alphav}
   \varphi-\alpha_m
 -\left(c+\widehat{v}_l\right) 
  \left(\frac{\partial{\cal T}}{\partial x_l}
       +\frac{\varphi}{(c+\widehat{v}_l)}\right) =0
\end{equation}

\noindent or, the equation for $\alpha_m$ in terms of the gradient of ${\cal T}$:

\begin{equation}\label{finalalpham}
  \alpha_m=-\left(1+\frac{\widehat{v}_l}{c}\right) 
	   c\frac{\partial{\cal T}}{\partial x_l}
         \equiv \left(1+\frac{\widehat{v}_l}{c}\right)\alpha .  
\end{equation}

\noindent This is the relationship between the bending angle and the
gravitational time delay at the lens plane when the lens is moving. It
represents a correction of a factor of $(1+v/c)$ with respect to the bending
angle for the same deflector at rest. This is in agreement with aberration
\cite{aberr}, according to which the angles made by lightrays as observed in a
moving frame are corrected by a factor of $(1+v/c)$ with respect to the angles
made by the same lightrays as observed in the rest frame. In our case both the
directions of incoming and outgoing lightrays are corrected by the same
factor, and consequently their deviation $\alpha $ is corrected by the same
factor.

The lens equation 

\begin{equation}\label{lensprelim}
	x=x_l+(y-y_l(\tau_l+{\cal T}))\tan(\varphi-\alpha_m),  
\end{equation}

\noindent is obtained from by eliminating $\tau-\tau_l$ from (\ref{vsource}).
We are interested in the small angle regime so, neglecting higher order terms,
we have

\begin{equation}\label{lensprelim2}
	x=x_l+(y-\widehat{y}_l)(\varphi-\alpha_m), 
\end{equation}

\noindent where we have used the assumption that ${\cal T}$ is small (of the
same order as $\varphi$ and $\alpha_m)$ and introduced the notation
$\widehat{y}_l\equiv y_l(\tau_l)$. From the ratio of (\ref{r=01}) over
(\ref{r=02}) and for small $\varphi$, we have

\begin{equation} \label{phixlyl}
   \varphi = \frac{x_l}{\widehat{y}_l} 
\end{equation}

\noindent which leads, via Eq.~(\ref{lensprelim2}), to

\begin{equation}\label{lensalmost}
	x = \frac{y}{\widehat{y}_l}x_l-(y-\widehat{y}_l)\alpha_m
\end{equation}

\noindent or 

\begin{equation}
   \frac{x}{y} = \frac{x_l}{\widehat{y}_l}
                -\frac{y-\widehat{y}_l}{y}\alpha_m
\end{equation}

\noindent Since in flat space the coordinate distance is equal to the
angular diameter distance, we can use the substitutions $y=D_s$, 
$\widehat{y}_l=\widehat{D}_l$ and $y-\widehat{y}_l=\widehat{D}_{ls}$ to denote
the distance to the source, the distance to the lens and the distance
between the lens plane and the source plane at the time the wavefront passes
by the lens plane. With these substitutions the lens equation becomes 

\begin{equation}\label{LE2}
   \frac{x}{D_s} = \frac{x_l}{\widehat{D}_l}
    		  -\frac{\widehat{D}_{ls}}{D_s}\alpha_m  
\end{equation}

Thus the changes in the lens equation from lens motion amount to a factor of 
$(1+\widehat{v}_l/c)$ in the bending angle (from Eq.~(\ref{finalalpham}))
and the evaluation of the instantaneous position of the lens plane at the
average time that the wavefront reaches it.

The time delay can be obtained by solving for $\tau$ in (\ref{ysource}) and
subtracting the travel time of a ray in the absence of the lens. We have thus

\begin{equation}
	\Delta\tau 
   =	\frac{1}{c}\frac{y-y_l(\tau_l+{\cal T})}
		        {\cos(\varphi-\alpha)}
	+\tau_l+{\cal T}-\frac{1}{c}\sqrt{x^2+y^2} 
\end{equation}

From Eq.~(\ref{LE2}) and Eq.~(\ref{r=02}), for small $\varphi$,

\begin{equation}\label{approxtaul}
   \tau_l = c^{-1}\widehat{y}_l(1+\frac12\varphi^2)
\end{equation}

\noindent but {\it keeping up to second-order terms}, we find the time delay
equation corrected for a moving lens;

\begin{equation}\label{deltat_1}
   	\Delta\tau 
   =	(1+\widehat{v}_l/c){\cal T}
	+\frac{\widehat{D}_l\widehat{D}_{ls}}
	      {2cD_s}\alpha^2.  
\end{equation}

\section{Cosmological Thin Lenses}


In sections IV and V, we derived the implicit Fermat potentials for lensing
by static and moving thin lenses where the lens plane separated two regions
of flat spacetimes. In this section, we expand these results to consider
lensing in spacetimes that (apart from the lens) are described by
Friedman-Robertson-Walker (FRW) metrics. Our approach is to utilize the fact
that all the FRW spacetimes are conformally flat to rescale physical
quantities in the flat lensing case into physical quantities for observers
living in a FRW universe. We begin by noting that there is, for each FRW
spacetime ($k=\pm 1,0$), a coordinate transformation between the ``natural''
FRW coordinates $(\tau,\chi,\theta,\phi)=x_F^a$ to ``Minkowski''
coordinates $(t,x,y,z)=x_M^a$,
 
\begin{equation}\label{coordinate}
	x_M^a = x_M^a(x_F^a),  
\end{equation}

\noindent such that the metric can be written as 

\begin{equation}\label{metric}
  d\widetilde{s}^2
= d\tau^2-a^2(\tau)dl_k^2(\chi,\theta,\phi)
=\Omega^2(dt^2-dx^2-dy^2-dz^2).  
\end{equation}

\noindent Here, $dl_k^2$ is the metric of a constant curvature three-space
(sphere for $k=+1$, hyperboloid for $k=-1$, or flat space for $k=0$). In
general, the conformal transformation is such that

\begin{equation}\label{omega}
	\Omega = f_k(x^a)a(\tau ),  
\end{equation}

\noindent for some function $f_k$. Because conformal transformations preserve
lightrays, the implicit Fermat potentials for lensing in our previous sections
are implicit Fermat potentials for the cosmological spacetime and can be
expressed in terms of either coordinate system, $x_F^a$ or $x_M^a$, via
Eq.~(\ref{coordinate}).

{\it Often, when a symbol is used for a quantity described in the Minkowski
space and we want to distinguish it from the same quantity in the cosmological
space we will indicate that by a twiddle, i.e., $A$ versus $\widetilde{A}$.
Also, often when we are refering to a quantity associated with a moving lens we
will indicate that by a ``hat'', e.g., the $y_l$ position of the moving lens
becomes $y_l(\tau)=\widehat{y}_l$.}

Several subtle issues arise from the fact that, in general, a fixed space point
$(\chi,\theta,\phi)=$constant moving with the cosmological flow in the $k=\pm 1$
cases, has in the associated Minkowski space a coordinate velocity; thus
co-moving lenses in the FRW coordinates are modeled by lenses that move, i.e.,
have a coordinate velocity, in the Minkowski space. This, in turn, leads to the
consideration of aberration affects in the bending angle and source angles which
then influence angular-diameter distances. A second subtle issue is the
relationship between the gravitational time delay as computed in the FRW spaces
compared to those computed in the associated Minkowski space. We point out that
the time delays are not simply conformally related; the cosmological time delay
must be calculated independently from the Minkowski delays. This issue will be
further explored later in this Section.

We consider the universe to be two sections of a FRW spacetime appropriately
matched at a lens plane. The matter distribution in the lens plane
determines two related gravitational time delay functions: the (Minkowski) 
{\it coordinate time} delay, ${\cal T}$, and the related
cosmological proper time delay, ${\cal T}_C$, with

\begin{equation}\label{proper}
  {\cal T}_C=\Omega_l{\cal T}  
\end{equation}

\noindent where $\Omega_l$ is the conformal factor evaluated at the lens at the
time of arrival of the ray. The relevant functions depend on the Minkowski
coordinates in the lens plane, $x_l.$ We interpret ${\cal T}_C(x_l)$ as the
cosmological proper time measured by a person just past the lens (on the
observer's side) between a lightray emitted at the source at time $\tau_s$ that
passed through the lens and one emitted at the same  $\tau_s$ that was not
influenced by the lens.

As we saw for the moving lens in Sec.V, our implicit Fermat potential, in 
2+1 dimensions, is given by (\ref{vsource}) with the
envelope condition yielding

\begin{equation} \label{alpham}
	  \alpha_m 
       = -(c+\hat{v}_l)\frac{\partial{\cal T}}{\partial x_l} \\
  \equiv  (1+\hat{v}_l/c)\alpha,  	     
\end{equation}

\noindent where $\hat{v}_l$ is the average speed of the lens during the time
the lightray is delayed and $\alpha_m$ is the bending angle observed by a
stationary observer with a moving lens and $\alpha$ the bending angle for an
observer comoving with the lens. Our lens equation (in the small angle
approximation) is simply

\begin{equation}\label{prelens}
	x=\varphi y-(y-\hat{y}_l)\alpha_m,  
\end{equation}

\noindent where $(x,y)$ are the source position, $\hat{y}_l$ is the position of
the lens plane when the lightray leaves the lens, and $\varphi$ is the
observation angle.

We consider how to introduce some (modified) angular-diameter distance
\textbf{s} for lenses that are moving in a Minkowski spacetime. Here, we
interpret the source coordinate, $x$, as the {\it metric distance}, 
$\Delta_s$, between the source (in the source plane) and the optical-axis,
(i.e. $x\equiv\Delta_s$). We then convert the coordinate distances, $y$, and
$y-\hat{y}_l$ into angular-diameter distances, $D_s$ and $\widehat{D}_{ls}$, as
follows. Let $\beta $ be the unlensed angle that the source at $x$ subtends at
the observer and let $\gamma $ be the angle that $x$ subtends at the lens plane
(when the lightray leaves the lens plane). Then

\begin{mathletters}
\begin{eqnarray}
   y &\equiv & D_s = \frac{x}{\beta}
		   =\frac{\Delta_s}{\beta}, \\
   y-\hat{y}_l &\equiv & \widehat{D}_{ls} = \frac{x}{\gamma}
				          = \frac{\Delta_s}{\gamma}.
\end{eqnarray}
\end{mathletters}

\noindent Eliminating $y$ and $y-\hat{y}_l$ from (\ref{prelens}), the lens
equation becomes

\begin{equation}
   \beta = \varphi -\frac{\widehat{D}_{ls}}{D_s}\alpha_m.  \label{lens1}
\end{equation}

This expression for the lens equation with a moving lens is in terms of
distances and angles relative to the frame of a stationary observer in the
Minkowski space. As we saw earlier, by the aberration (of light) equation,
the bending angle seen by the stationary observer, $\alpha_m$, is related
to the bending angle $\alpha$ seen in the frame comoving with the lens, by
Eq.~(\ref{alpham}),

\begin{equation}
\alpha_m=\alpha (1+\frac{\hat{v}_l}{c}),  \label{aber}
\end{equation}

\noindent so that Eq.~(\ref{lens1}) becomes

\begin{equation}\label{lensa}
   \beta =\varphi -\frac{\widehat{D}_{ls}}{D_s}\alpha_m
         =\varphi -\frac{\widehat{D}_{ls}}{D_s}\alpha 
	  (1+\frac{\hat{v}_l}{c})  
\end{equation}

\noindent with $\widehat{D}_{ls}$ the time-dependent angular-diameter distance
of the lens to the source {\it as seen by a stationary observer}. On the other
hand, if we wish to use (as we will in a moment) the angular-diameter distance
measured from the lens to the source {\it in the comoving lens frame},
$D_{ls}^{(co)}$, we have that

\begin{equation}
	      D_{ls}^{(co)}
   =\frac{x}{\gamma^{(co)}}
   =\frac{\Delta_s}{\gamma^{(co)}}, 
\end{equation}

\noindent where $\gamma^{(co)}$ is the angle that $x$ subtends according to
an observer comoving with the lens plane. Using the aberration equation,
Eq.~(\ref{aber}), for the angle $\gamma$, we have

\begin{equation}\label{dco}
   \widehat{D}_{ls} = D_{ls}^{(co)}\frac{1}{(1+\frac{v}{c})}.  
\end{equation}

\noindent The (moving) lens equation becomes

\begin{equation}\label{lens2}
   \beta = \varphi -\frac{D_{ls}^{(co)}}{D_s}\alpha .  
\end{equation}

Note that neither $D_{ls}^{(co)}$ or $\widehat{D}_{ls}$ are directly
observable quantities. The issue of which to use depends on the physical
situation that is being addressed. If we utilize the coordinate
transformation, Eq.~(\ref{coordinate}), that relates the Minkowski
coordinates to the ``natural'' FRW coordinates, the moving lens in the
Minkowski background becomes a stationary lens in the conformally related
cosmological spacetime. Hence, $D_{ls}^{(co)}$ should be used in the lens
equation in the case where our interest lies in the FRW space.

The just completed discussion pertained to the Minkowski background; we now
transform the lens equation, Eq.~(\ref{lens2}), to the cosmological background.
We assume that the observer's worldline is given by
$(\tau,\chi=0,\theta=0,\phi=0)$ in the standard FRW coordinates. In any of the
three FRW models, such an observer is stationary at $(t,r=0,\theta=0,\phi=0)$.
However, the lens and source, located along the worldlines of constant 
$(\chi,\theta,\phi)$ will be ``moving'' in the Minkowski coordinates.  (The
motion of the source in the Minkowski coordinates plays no important role.) In
the cosmological space, the metric distance between the source location and the
optical axis (in the source plane) is given by $\widetilde{\Delta}_s=\Omega
(x_s^a)\Delta_s$ where the conformal factor is evaluated at the source
location when the lightray leaves the source. Using

\begin{mathletters}
\begin{eqnarray}
\widetilde{D}_{s} &=&\frac{\widetilde{\Delta}_s}{\beta}
		   =\Omega(x_s^a)D_s  				\\
\widetilde{D}_{ls}&=&\frac{\widetilde{\Delta}_s}{\gamma^{(co)}}
	           =\Omega(x_s^a)D_{ls}^{(co)}
\end{eqnarray}
\end{mathletters}

\noindent for the cosmological angular-diameter distances, our lens equation
immediately becomes 

\begin{equation}\label{lens-fin2}
  \beta = \varphi -\frac{\widetilde{D}_{ls}}{\widetilde{D}_s}\alpha ,
\end{equation}

\noindent which is identical (in form) to the flat-space lens equation with a
stationary lens and to the conventional cosmological lens
equation~\cite{EFS,Petters}.

We now turn from the cosmological lens equation to the cosmological time of
arrival equation. In Section~V, we derived an expression for the ``time
delay'' at the observer between the true ``lensed'' path and the
``unlensed'' path in a Minkowski model with a moving lens, namely,

\begin{equation}\label{arrival}
  	\Delta\tau 
   =	(1+\hat{v}_l/c){{\cal T}}
	+\frac{\widehat{D}_l}{2c}
	 \frac{\widehat{D}_{ls}}{D_s}\alpha_m^2.  
\end{equation}

\noindent Using the transformation properties of $\alpha_m$ and
$\widehat{D}_{ls}$ just described, Eq.~(\ref{arrival}) has the form

\begin{equation}\label{time1}
	\Delta \tau 
   =	(1+\hat{v}_l/c)
	\left({\cal T}+\frac{D_l}{2c}
		       \frac{D_{ls}^{(co)}}{D_s}\alpha^2\right) ,  
\end{equation}

\noindent where $\alpha$ is the bending angle seen in the comoving frame and
${\cal T}$ is the {\it Minkowski coordinate time-delay} measured along a
worldline just to the observer's side of the lens between the arrival time for a
lensed and unlensed ray. To transform Eq.~(\ref{time1}) to cosmological
variables, we first note that 

\begin{equation}
   \frac{D_{ls}^{(co)}}{D_s}
  =\frac{\widetilde{D}_{ls}}{\widetilde{D}_s} 
\end{equation}

\noindent where the ratio on the right side refers to cosmological
angular-diameter distances. Next, we see that if $\Delta_l$ is the
Minkowski metric distance from the optical axis to the ``image location'' in
the lens plane and $\varphi$ is the angle this distance subtends at the
observer we have 

\begin{equation}
   \widetilde{D}_{l}
  =\frac{\Omega(x_l^a)\Delta_l}{\varphi}
  =\Omega(x_l^a)\widehat{D}_l, 
\end{equation}

\noindent which relates the angular-diameter distance from observer to the lens
in the FRW cosmology to the conformally related ``Minkowski'' angular-diameter
distance. (Here, the conformal factor is evaluated at the location of the lens
at the time the lightray leaves the lens.) If we multiply both sides of
Eq.~(\ref{time1}) by $\Omega_0\Omega_l$ where $\Omega_0$ and $\Omega_l$ are
respectively the conformal factor evaluated at the observer at the observation
time and at the lens at the time the light ray leaves the lens, we have 

\begin{equation}\label{time2}
	\Delta\widetilde{\tau}_0
       =(1+\hat{v}_l/c)\frac{\Omega_0}{\Omega_l}
	\left({\cal T}_C
	     +\frac{\widetilde{D}_l}{2c}
	      \frac{\widetilde{D}_{ls}}{\widetilde{D}_s}\alpha^2\right),  
\end{equation}

\noindent where we have used

\begin{mathletters}
\begin{eqnarray}
	{\cal T}_{C}             &=&\Omega_l{\cal T}  		\label{tauC} \\
	\Delta\widetilde{\tau}_0 &=&\Omega_0\Delta\tau  \label{deltatautwid}
\end{eqnarray}
\end{mathletters}

\noindent the latter being the cosmological proper time between reception of
the two rays at the observer.

From the relation 

\begin{equation}\label{appendix}
   (1+\hat{v}_l/c)\frac{\Omega_0}{\Omega_l} \approx \frac{a_0}{a_l}
					    \equiv   1+z,  
\end{equation}

\noindent which is derived in the appendix for small coordinate velocities,
$\hat{v}_l/c$, in the associated Minkowski space, from Eq.~(\ref{time2}), our
final cosmological time delay equation is

\begin{equation}\label{time3}
   \Delta \widetilde{\tau}_0
  =(1+z)\left({\cal T}_C
	      +\frac{\widetilde{D}_l}{2c}
	       \frac{\widetilde{D}_{ls}}{\widetilde{D}_s}\alpha^2\right).
\end{equation}

Our last issue is to describe how the cosmological proper time delay,
${\cal~T}_C$ is to be computed. To compute ${\cal T}_C$, and, by association,
the Minkowksi coordinate time delay, ${\cal T}=\Omega_l^{-1}{\cal T}_C$, we
first consider the Newtonian potential 

\begin{equation}\label{U}
	U_C
       =-\int\frac{dm'}{|\vec{x}-\vec{x}'|_C},  
\end{equation}

\noindent where $dm'=\rho_C(x')d^3x'$ is the ``observable'' mass element in the
cosmological space, and {\it the norm in the denominator is taken in the
cosmological metric}. To find the cosmological gravitational time delay, one
integrates the Newtonian potential over the path from the source to the
observer,

\begin{equation}\label{T}
{\cal T}_C=\frac{-2}{c^3}\int U_C dl_C,  
\end{equation}

\noindent where $dl_C$ is {\it the distance element along the path in the
cosmology}. In the standard thin-lens approximation, it is assumed that 
$\rho_C(x')$ is non-zero only in a thin region perpendicular to the optical
axis. Thus, when computing the time delay from the integral in Eq.~(\ref{T}),
one need only consider the region along the trajectory that is very close to the
lens plane. First, we point out that under conformal rescaling, since $dm'$ is
conformally invariant, that

\begin{equation}\label{UC}
	  U_C
        = -\int\frac{dm'}{|\vec{x}-\vec{x}'|_C}
  \approx -\int\frac{dm'}{\Omega_l|\vec{x}-\vec{x}'|_M}
        = \Omega_l^{-1}U_M  
\end{equation}

\noindent where $|\vec{x}-\vec{x}'|_M=$ is the metric distance between two
points in the lens plane taken in the conformally related Minkowski metric.
Because $U_C$ only has support in the vicinity of the lens and $dl_C=\Omega
dl_M$, where $dl_M$ is the distance element in a Minkowski spacetime, we have

\begin{equation}\label{Tm}
	{\cal T}_C
      = \frac{-2}{c^3}\int U_C dl_C
      = \frac{-2}{c^3}\int U_Md l_M
      = {\cal T}_M.  
\end{equation}

There is an apparent conundrum here. ${\cal T}_C$ is the physical proper-time
delay which {\it appears to be the same} as the Minkowski space proper-time
delay for the same physical situation, i.e., arising from potentials

\begin{mathletters}
\begin{eqnarray}
	U_C &=&-\int\frac{dm'}{|\vec{x}-\vec{x}'|_C} \\
	U_M &=&-\int\frac{dm'}{|\vec{x}-\vec{x}'|_M}
\end{eqnarray}
\end{mathletters}

with the same $dm'$ in both cases. The conundrum is resolved by
noting that though the $dm'$ are the same the density functions are
different.

We compute the mass densities by introducing {\it physical coordinates}
at the lens (local Lorentzian coordinates), so that the cosmological metric 
{\it at the lens} takes the form

\begin{equation}\label{Cflat}
    ds^2
   =(dt_C^2-dx_C^2-dy_C^2-dz_C^2)
   =\Omega_l^2(dt^2-dx^2-dy^2-dz^2)  
\end{equation}

\noindent with

\begin{equation}\label{xC}
	x_C^a=\Omega_lx^a.  
\end{equation}

\noindent Using

\begin{equation}\label{rho}
	dm'
      =\rho_C(x_C^a)d^3x_C
      =\rho_C(\Omega_lx^a)\Omega_l^3d^3x
      =\rho_M(x^a)d^3x  
\end{equation}

\noindent we have the relationship between the physical density, $\rho_C$,
and the {\it equivalent fictitious} Minkowski space density $\rho_M(x^a)$,

\begin{equation}
   \rho_M(x^a)
  =\rho_C(\Omega_lx^a)\Omega_l^3.
\end{equation}

Returning to the issue of the cosmological lens equation,
Eq.~(\ref{lens-fin2}), and associated bending angle, Eq.~(\ref{alpham}), we
see that the present results concerning the cosmological time-delay agree with
the earlier results, since from Eqs.~(\ref{tauC}), (\ref{xC}) and
(\ref{alpham}), we have

\begin{equation}\label{alphaC}
    \alpha 
   =-c\frac{\partial{\cal T}}{\partial x_l}
   =-c\frac{\Omega_l\partial{\cal T}}{\Omega_l\partial x_l}
   =-c\frac{\partial{\cal T}_C}{\partial x_{Cl}},  
\end{equation}

\noindent the cosmological bending angle.

All our results obtained from the envelope construction now agree with the
standard results appearing in the published literature.  In addition, it is
clear that the method can be used to extend the results to the case of a lens
plane that has a peculiar motion above the Hubble flow by letting the velocity
of the lens plane in the underlaying flat space be arbitrary. 

As a final comment here, we point out that this method of obtaining
the FRW lens equation and time delay equation by conformal rescaling of an
associated Minkowski space can be extended, largely unchanged, to any
spacetime that is conformally related to Minkowski space. It is not clear
whether this observation has any physical application.

\section{Concluding Remarks and Outlook}


We have introduced a novel approach to gravitational lensing built on a
variational principle that is analogous to, but distinctly different from, the
conventional version of Fermat's principle.

Our approach is to find the envelope of an appropriate family of null surfaces
containing the observer's location. We have shown how this approach reproduces
the astrophysical scenario of static thin lenses as an illustration. More
interestingly, though, we have been able to use this approach to obtain the
correction to first order in $v/c$ that the motion of the deflector imposes on
the bending angle in the approximation of thin lenses. This calculation can
alternatively be done by direct integration of the null geodesics in
time-dependent spacetime perturbations off flat space. In this respect, our
result agrees with such a direct calculation of the bending angle by moving
thin lenses carried out by Pyne and Birkinshaw~\cite{pyne1}, whereas it differs
from others~\cite{Italians} by an overall sign and a factor of 2. Work is in
progress to clarify this difference and will be reported elsewhere.  Notice
that a redshift of 0.001 might conceivably bring this effect into the
observable regime in the future.

Because in the thin-lens case in cosmology our approach is based entirely on
the underlying conformally flat space, we also have an alternative derivation
of the lens equation in cosmology, which does not seem to have been exploited
in the literature so far.  We feel our derivation clarifies some points that
remain obscure in the presentations of the lens mapping in cosmology that
appear in \cite{EFS} and \cite{Petters}. On the other hand, the reader
interested in a very complete derivation of the lens mapping in cosmology by
integration of the null geodesics will definitely benefit from the excellent
article by Pyne and Birkinshaw~\cite{pyne2}.   

Throughout this work, we have assumed that a metric is given that represents
the structure of the deflector or lens. In this sense, we have kept ourselves
within the kinematics of the implicit Fermat potential. We have not addressed
the issue of the dynamics of the Fermat potential, namely: how the implicit
Fermat potential is directly affected by the structure of the deflectors. The
resolution of this important question in principle involves two steps: to solve
the Einstein equations for the metric, and then to use the metric as a given
source into the eikonal equation for the implicit Fermat potential. By
contrast, in the standard thin-lens scenario, the Fermat potential is obtained
directly from the surface mass distribution by means of a 2-dimensional
equation of the Poisson type. It is reasonable to ask whether an analogous
scheme to obtain field equations for the implicit Fermat potential directly in
terms of the structure of the deflector exists in cases other than the
thin-lens scenario. This question will be discussed elsewhere.

\acknowledgments

We are indebted to Juergen Ehlers for a very enlightening discussion. This
work was supported by the NSF under grants No. PHY-0088951 and No.
PHY-0070624.

\appendix
\section*{}

In the main text we quoted the approximate equation,

\begin{equation}\label{result}
   \frac{a_{0}}{\Omega _{0}}\approx \frac{a_{l}}{\Omega _{l}}(1+v_{l}/c).
\end{equation}

\noindent which relates the ratio of conformal factors $\Omega$ to the
cosmological scale factor $a$, at two different points, the observer and the
lens, that are connected by a null geodesic. The quantity $v_l$ is the velocity
of the lens in the associated Minkowski coordinates evaluated as the ray leaves
the lens. 

For the case of $k=0$ this result is trivially true since then $a=\Omega$
and $v_l=0.$ For ease of presentation we will only give the proof for the 
$k=-1$ spacetime, but the calculations are exactly paralled in the case of 
$k=+1$ and the result is exactly the same. We start by recalling that we have
the metric in the two forms

\begin{mathletters}
\begin{eqnarray}
      d\widetilde{s}^2 
  &=& a^2(d\eta^2-d\chi^2 
	  -\sinh^2\chi(d\theta^2+\sin^2\theta d\varphi^2)) \\
  &=& \Omega^2(c^2dt^2-dr^2-r^2(d\theta^2+\sin^2\theta d\varphi^2))
\end{eqnarray}
\end{mathletters}

\noindent with Eq.~(\ref{coordinate}) explicitly given by

\begin{mathletters}\label{coordinate2}
\begin{eqnarray}
   ct &=&\frac{\sinh\eta }{\cosh\eta +\cosh\chi}   \\
    r &=&\frac{\sinh\chi}{\cosh\eta +\cosh\chi}  
\end{eqnarray}
\end{mathletters}

\noindent and
 
\begin{equation}
  \Omega^2 = \frac{4a^2}{(1-(ct+r)^2)(1-(ct-r)^2)},
\end{equation}

\noindent or, from Eq.~(\ref{coordinate2}),

\begin{equation} \label{omegaovera}
   \frac{\Omega}{a} = \cosh\chi +\cosh\eta . 
\end{equation}

\noindent We are interested in a lens at some fixed value of
$\theta,\varphi$ and $\chi=\chi_l$, sending a light signal to the observer
located at $\chi=0$.  The lens and arrival times are $\eta_l$ and $\eta_0$,
respectively, and we have

\begin{equation}\label{arrivaltime}
   \eta_0-\eta_l=\chi_l.  
\end{equation}

\noindent Using Eqs.~(\ref{omegaovera}) and (\ref{arrivaltime}) we can
construct

\begin{equation}
   \frac{\Omega_l}{a_l}
  =\frac{\Omega_0}{a_0}
   \frac{\cosh\chi_l+\cosh(\chi_l-\eta_0)}{1+\cosh\eta_0}
\end{equation}

\noindent or, for small $\chi_l$,

\begin{equation}\label{preresult*}
        \frac{\Omega_l}{a_l}
\approx \frac{\Omega_0}{a_0}
        \left(1-\frac{\chi_l\sinh\eta_0}
	             {1+\cosh\eta_0}\right).  
\end{equation}

By calculating the velocity from

\begin{equation}
    \frac{dr}{dt} 
  = \frac{\frac{\partial r}{\partial\eta}|_{\chi}}
         {\frac{\partial t}{\partial\eta}|_{\chi}}
\end{equation}

\noindent we have

\begin{equation}
        \frac{v}{c}
\equiv -\frac{\sinh\chi\sinh\eta}{1+\cosh\chi\cosh\eta}
\end{equation}

\noindent We thus see that a FWR lens plane approaches the observer in the
conformally related Minkowski space. Evaluating at the lens time we have

\begin{equation}
   \frac{v_l}{c}
  =\frac{\sinh\chi_l\sinh(\chi_l-\eta_0)}
        {1+\cosh\chi_l\cosh(\chi_l-\eta_0)}
\end{equation}

\noindent and for small values of $\chi_l$ 

\begin{equation}\label{vl*}
         \frac{v_l}{c}
\approx -\frac{\chi_l\sinh\eta_0}{1+\cosh\eta_0}.
\end{equation}

\noindent Using Eq.~(\ref{vl*}) with (\ref{preresult*}), we have our result,
Eq.~(\ref{result}).

\newpage

\end{document}